\documentclass[twocolumn,superscriptaddress,floatfix,noeprint]{revtex4-2}
\usepackage{graphicx,tabularx,makecell}
\usepackage{amsmath,amsfonts,amssymb}
\usepackage{mathtools}
\usepackage{hhline}
\usepackage{bm,dsfont}
\usepackage[dvipsnames]{xcolor}
\usepackage[colorlinks=true, citecolor=Green, linkcolor=BrickRed, urlcolor=NavyBlue]{hyperref}
\usepackage[all]{hypcap}
\usepackage[T5,T1]{fontenc}

\graphicspath{{figures/}}

\newcommand{\G}{\bm{\mathcal{G}}}

\begin{document}

\title{Roses in the Nonperturbative Current Response of Artificial Crystals}

\author{Christophe De Beule}
\email{cdebeule@sas.upenn.edu}
\affiliation{Department of Physics and Astronomy, University of Pennsylvania, Philadelphia PA 19104}
\affiliation{Department of Physics and Materials Science, University of Luxembourg, L-1511 Luxembourg, Luxembourg}
\author{{\fontencoding{T5}\selectfont V\~o Ti\'\ecircumflex{}n Phong}}
\affiliation{Department of Physics and Astronomy, University of Pennsylvania, Philadelphia PA 19104}
\author{E. J. Mele}
\affiliation{Department of Physics and Astronomy, University of Pennsylvania, Philadelphia PA 19104}
\date{\today}

\begin{abstract}
In two-dimensional artificial crystals with large real-space periodicity, the nonlinear current response to a large applied electric field can feature a strong angular dependence, which encodes information about the band dispersion and Berry curvature of isolated electronic Bloch minibands. Within the relaxation-time approximation, we obtain analytic expressions up to infinite order in the driving field for the current in a band-projected theory with time-reversal and trigonal symmetry. For a fixed field strength, the dependence of the current on the direction of the applied field is given by rose curves whose petal structure is symmetry constrained and is obtained from an expansion in real-space translation vectors. We illustrate our theory with calculations on periodically-buckled graphene and twisted double bilayer graphene, wherein the discussed physics can be accessed at experimentally-relevant field strengths.
\end{abstract}

\maketitle

\begin{figure}
    \centering
    \includegraphics[width=\linewidth]{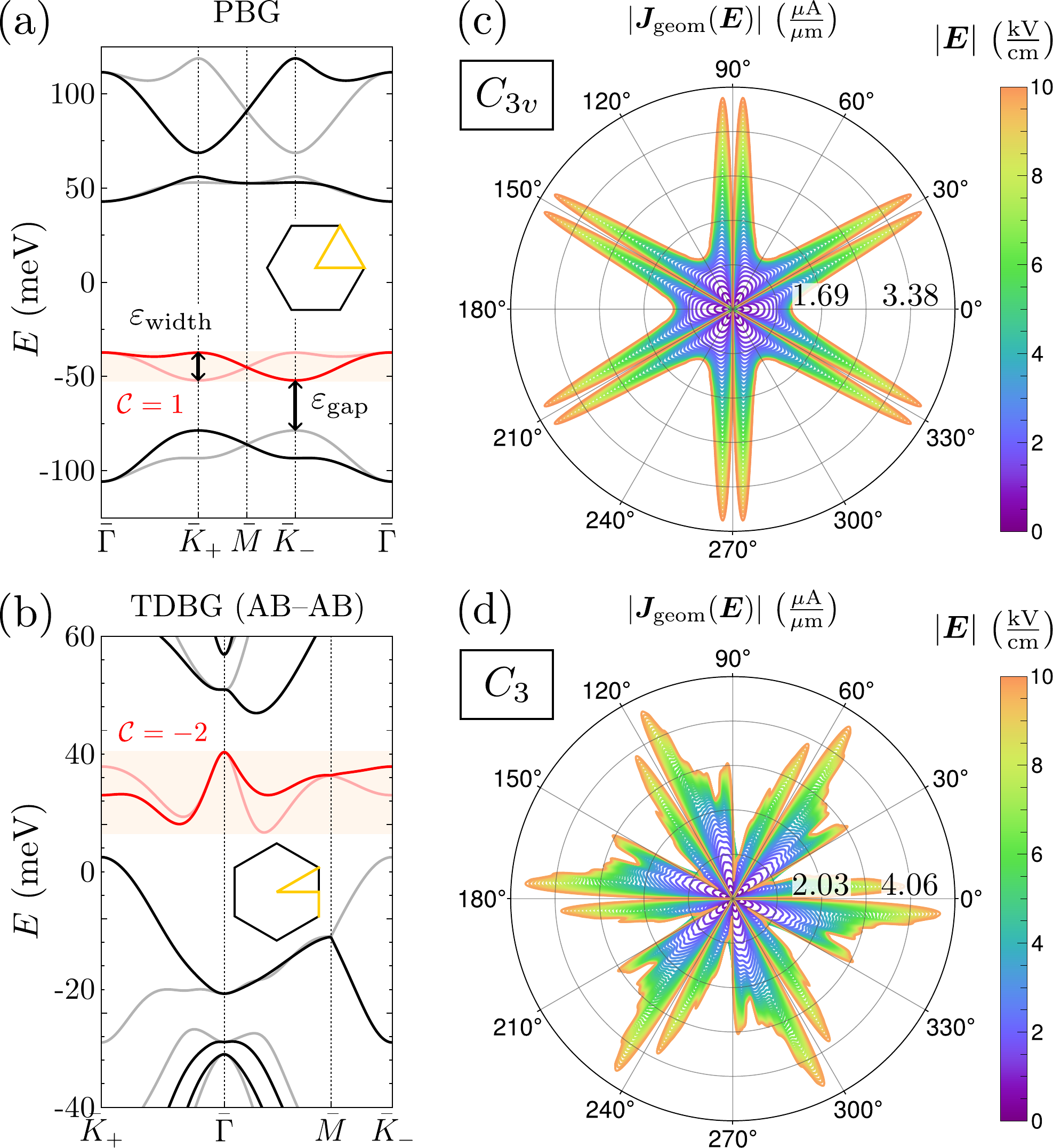}
    \caption{(left) Energy bands for (a) periodically-buckled graphene and (b) AB--AB twisted double bilayer graphene, along high-symmetry lines in the superlattice Brillouin zone. Dark/light bands correspond to valley $K_+$/$K_-$ and the highest valence [lowest conduction] band is shown in red in (a) [(b)] with valley Chern number $\mathcal C$. (right) Geometric current of the highlighted bands at 0.4 filling and $T=5$~K for PBG (c) and TDBG (d). Parameters for PBG are: $L/l_0 = 6$, $\mathcal V_0 \approx 26\;\text{meV}$, and $3\phi/\pi=-0.155$ with $\varepsilon_\text{gap}^2 / \varepsilon_\text{width} = 48 \; \text{meV}$; and for TDBG: $\vartheta = 1.44^\circ$ and $U=56.5\;\text{meV}$ and others from Ref.\ \cite{Koshino2019} with $\varepsilon_\text{gap}^2 / \varepsilon_\text{width} = 4.1\; \text{meV}$.}
    \label{fig:fig1}
\end{figure}

In two-dimensional (2D) crystals, rotation symmetries about the axis perpendicular to the 2D plane require that the DC current response $\bm J(\bm E)$ to a constant uniform electric field $\bm E$ is isotropic to first order in the field strength. Given a crystal symmetry $\mathcal S$, the current obeys $\bm J(\mathcal S \bm E) = \mathcal S \bm J(\bm E)$ and unless there are very few symmetries, anisotropies generally occur at higher order \cite{wu2017giant}. Nevertheless, the anisotropy in the current can be a valuable tool for probing the energetic and geometric properties of electron bands. In systems with atomic-scale periodicity, the strong-field regime is not readily accessible because the required fields generally induce interband transitions, i.e., electric breakdown \cite{AshcroftMermin}, which mask the properties of an otherwise isolated band. However, in moir\'es \cite{Andrei2020,Andrei2021,Mak2022} and superlattice heterostructures \cite{Tsu}, such as periodically gated \cite{Forsythe2018} or strained \cite{Mao2020} systems, the spatial period of the lattice $L$ can be made large, of the order  $10 \; \text{nm}$. Hence, a nonperturbative regime, which we define as
\begin{equation}
    \omega_B \tau \gg 1, \qquad \omega_B = eEL / \hbar,
\end{equation}
with $\omega_B$ the Bloch frequency and $\tau$ the momentum-relaxation time, can be reached for realistic field strengths \cite{Fahimniya2021}. Taking $L = 10 \; \text{nm}$ and $\tau = 1 \; \text{ps}$ \cite{Dawlaty2008}, we find $E \gg 0.66 \; \text{kV/cm},$ which is experimentally feasible. Importantly, the strong-field regime can be realized in these systems well below the onset of electric breakdown, which we estimate as follows. We require $e E \Delta x \ll \varepsilon_\text{gap}$ with $\Delta x$ the uncertainty in the position of the electron, and thus $e E \ll \varepsilon_\text{gap} \hbar  \Delta v / |\partial^2 \varepsilon/\partial k^2|$. Assuming the curvature is largest near the band edge and that $\Delta v/v$ should be small in the semiclassical theory \cite{AshcroftMermin}, we find $eEL \ll \varepsilon_\text{gap}^2 L / \hbar v < \varepsilon_\text{gap}^2 / \varepsilon_\text{width}$ where $\varepsilon_\text{width}$ is the bandwidth. Since moir\'es and other artificial crystals can host spectrally isolated and narrow minibands, as shown in Fig.\ \ref{fig:fig1}(a) and (b) for periodically-buckled graphene (PBG) and twisted double bilayer graphene (TDBG), respectively, the right-hand side of this inequality can be made large. For example, taking $L = 10 \; \text{nm}$ and $\varepsilon_\text{gap}^2 / \varepsilon_\text{width} = 50 \; \text{meV}$, which we find can be realized in PBG, we obtain $E \ll 50 \; \text{kV/cm}$.

In this work, we investigate the nonperturbative current response in a band-projected theory. That is, we solve the semiclassical transport theory exactly to infinite order in the field strength, but assume that interband transitions are negligible, as outlined in the previous paragraph. We focus on trigonal systems with time-reversal ($\mathcal T$) symmetry for which $\mathcal C_{3z}$ rotation symmetry is conserved but $\mathcal C_{2z}$ rotation symmetry and inversion symmetry are broken, i.e., the point groups $C_3$, $C_{3v}$, $C_{3h}$, $D_3$, and $D_{3h}$ \cite{Dresselhaus}. These are the relevant point groups of many moir\'es and other 2D superlattices, such as those based on graphene \cite{Andrei2020,Mao2020} and transition-metal dichalcogenides \cite{Mak2022}. We first consider the weak-field limit and determine the lowest-order anisotropy in the currents from symmetry. There we find a nonlinear Hall response from the Berry curvature hexapole, since the lower order dipole response is forbidden by $\mathcal C_{3z}$ and odd powers are forbidden by $\mathcal T$. In contrast, the current originating from the band dispersion becomes anisotropic in the transverse response at fifth order while the longitudinal response only depends weakly on the field direction. Furthermore, by expanding in terms of coordination shells, we obtain analytic expressions for the current in terms of the \emph{real space} Fourier components of the band dispersion and the Berry curvature. This differs from the usual prescription where one expresses the current in terms of multipole moments at successive orders in the response \cite{Sodemann2015,Zhang2023}. Our results amount to resumming the multipoles to infinite order in the field strength, elucidating the dependence of the current on the field direction. Here we focus on the current instead of conductivities since the latter are harder to interpret at arbitrary order. For a fixed field strength, we find that the anisotropic current distributions take the form of \emph{rose curves} when plotted as a function of the field direction, as illustrated in Fig.\ \ref{fig:fig1}(c) and (d). While the longitudinal response has no petals, the main petal structure of the transverse response is determined from symmetry. These petals can fracture into an odd number of subpetals by breaking in-plane mirrors (or equivalently, out-of-plane $\pi$ rotations). In a recent work, two of the authors demonstrated that the geometric current originating from the Berry curvature plateaus and dominates in the strong-field limit \cite{Phong2022b}. Here, we obtain the exact plateau values which are strongly anisotropic. We finally apply our theory to graphene-based artificial crystals that host spectrally isolated and narrow minibands: periodically-buckled graphene and twisted double bilayer graphene.

\let\oldaddcontentsline\addcontentsline 
\renewcommand{\addcontentsline}[3]{} 
\section{Weak-Field Response}

It is illustrative to first consider the weak-field limit ($\omega_B \tau \ll 1$) to investigate how symmetry constrains the order at which anisotropy sets in. To this end, we write the current as $\bm J = \bm J^{(+)} + \bm J^{(-)}$ where $\bm J^{(\pm)}(\bm E) \equiv \tfrac{1}{2} \left[ \bm J(\bm E) \pm \bm J(-\bm E) \right]$ are, respectively, even and odd under field reversal ($\bm E \mapsto -\bm E$). Note that $\bm J^{(+)}$ vanishes if inversion or $\mathcal C_{2z}$ symmetry is conserved. In the presence of $\mathcal C_{3z}$, the currents can be expanded as 
\begin{align}
    J_x^{(-)} + i J_y^{(-)} & = a(E^2) E_+ + b E_-^5 + \mathcal O(E^7), \label{eq:jodd} \\
    J_x^{(+)} + i J_y^{(+)} & = c(E^2) E_-^2 + d E_+^4 + \mathcal O(E^6), \label{eq:jeven}
\end{align}
where $E_\pm = E_x \pm i E_y$, $a = a_0 + a_1 E^2 + a_2 E^4$, $c = c_0 + c_1 E^2$, and all other parameters are c-numbers. In particular, $a_0 = \sigma_L + i \sigma_H$ where $\sigma_L$ ($\sigma_H$) is the longitudinal (Hall) conductivity. Both sides in \eqref{eq:jeven} and \eqref{eq:jodd} transform as angular momentum $L_z = 1$ objects  which are conserved mod $3$ under $\mathcal C_{3z}$ symmetry. If mirror symmetry $\mathcal M_x$ ($x \mapsto -x$) is also conserved, one finds that $a$ and $b$ are real while $c$ and $d$ are imaginary, while $\mathcal M_y$ ($y \mapsto -y$) makes all parameters real. Time-reversal symmetry further requires that $a$ is real by Onsager reciprocity. Similar expressions were obtained for the odd current in the presence of $\mathcal C_{2z}$ or $\mathcal C_{4z}$, while $\mathcal C_{6z}$ results again in \eqref{eq:jodd}. These are given in the Supporting Information (SI, section I C).

The anisotropy in the current is most clearly expressed in terms of the longitudinal $J_\parallel \equiv \hat E \cdot \bm J$ and transverse $J_\perp \equiv ( \hat E \times \hat z ) \cdot \bm J$ current components with $\hat E = \left( \cos \theta, \sin \theta \right)$ the field direction. This is because $J_\parallel$ ($J_\perp$) transforms as a scalar (pseudoscalar) under a crystal symmetry. For example, for $C_{3v} = \left< \mathcal C_{3z} , \mathcal M_x \right>$ (or $D_3$) and $\mathcal T$ symmetry,
\begin{gather}
    J_\perp^{(+)} = 2\tilde c E^4 \cos(3 \theta), \\
    J_\parallel^{(-)} = aE + bE^5 \cos( 6\theta ), \quad J_\perp^{(-)} = 2 b E^5 \sin (6 \theta),
\end{gather}
at leading order with real-valued $\tilde c = ic_1$. Here we have assumed that the even current is purely transverse. Note that the projected currents gain an extra sign under field reversal. The lowest nonzero Hall effect thus originates from the Berry curvature hexapole in a system with $\mathcal C_{3z}$ and $\mathcal T$ symmetry \cite{Zhang2023,Leppenen2023}.

\section{Expansion in Coordination Shells}

To calculate the current response, we start from the equations of motion for electrons in a 2D crystal \cite{Chang1995,Sundaram1999}:
\begin{equation}
    \hbar \dot{\bm r}_{\bm k} = \nabla_{\bm k} \varepsilon_{\bm k} - \hbar \dot{\bm k} \times \Omega_{\bm k} \hat z, \qquad \hbar \dot{\bm k} = -e\bm E,
\end{equation}
where $\Omega_{\bm k} = -2 \, \text{Im} \left( \langle \partial u_{\bm k}/\partial k_x | \partial u_{\bm k}/\partial k_y \rangle_\text{cell} \right)$ is the Berry curvature with $u_{\bm k}(\bm r)$ the cell-periodic Bloch functions in periodic gauge \cite{Vanderbilt2018}, and  $\varepsilon_{\bm k}$ is the band dispersion. The steady-state current is given by $\bm J = -2e \int_\text{BZ} \frac{d^2\bm k}{(2\pi)^2} \, f_{\bm k} \dot{\bm r}_{\bm k} \equiv \bm J_\text{Bloch} + \bm J_\text{geom}$ with $f_{\bm k}$ the non-equilibrium distribution function, obtained from the Boltzmann transport equation. In the relaxation-time approximation, by resumming the solution to all orders in the electric field, we find
\begin{align}
    \bm J_\text{Bloch} & = \frac{2e}{V_c \hbar} \sum_{\bm R} \frac{i \bm R f_{\bm R}^0 \varepsilon_{-\bm R}}{1 - i e \tau \bm E \cdot \bm R / \hbar}, \label{eq:JBloch} \\
    \bm J_\text{geom} & = \left( \hat z \times \bm E \right) \frac{2e^2}{V_c \hbar} \sum_{\bm R} \frac{f_{\bm R}^0 \Omega_{-\bm R}}{1 - i e \tau \bm E \cdot \bm R / \hbar}, \label{eq:Jgeom}
\end{align}
where the sums run over lattice vectors, $V_c$ is the unit cell area, and $\varepsilon_{\bm R}$, $f^0_{\bm R}$, and $\Omega_{\bm R}$ are Fourier components with $f^0_{\bm k} = f^0( \varepsilon_{\bm k})$ the Fermi function. The factor two accounts for spin, as we assume spin-orbit coupling is weak throughout this work.

\begin{figure}
    \centering
    \includegraphics[width=\linewidth]{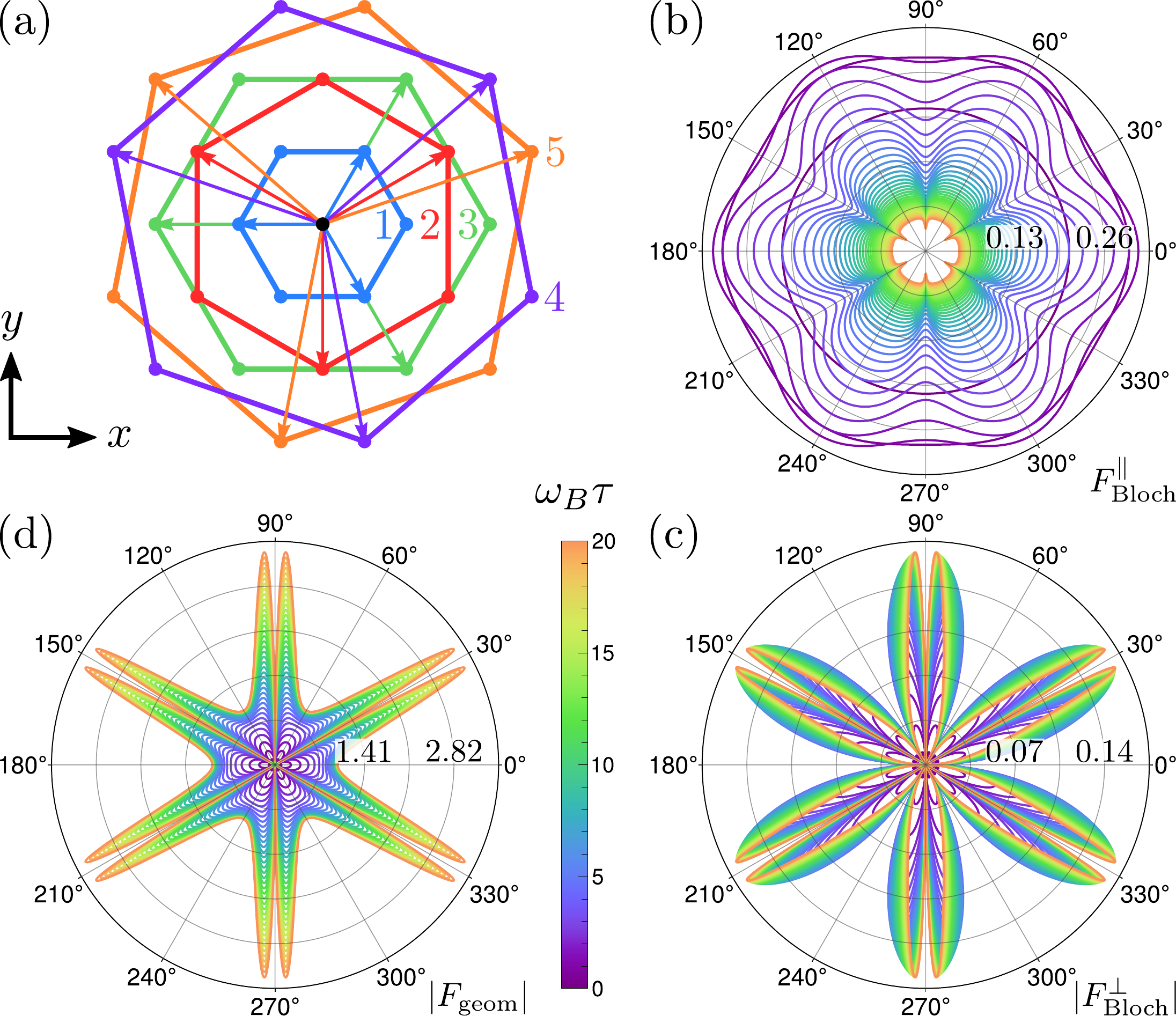}
    \caption{(a) First five coordination shells of the triangular lattice. (b)--(d) First-shell roses for the longitudinal (b), transverse Bloch (c), and geometric (d) current. The color scale gives the value of $\omega_B \tau = e\tau EL/\hbar$.}
    \label{fig:fig2}
\end{figure}

We now consider a Chern trivial band, separated in energy from other bands, with $C_{3v} = \left< \mathcal C_{3z} , \mathcal M_x \right>$ (or $D_3$) and $\mathcal T$ symmetry. Expanding in coordination shells:
\begin{equation}
    \varepsilon_{\bm k} = \sum_{j,n} \varepsilon_j \cos ( \bm k \cdot \bm L_n^{(j)} ), \quad
    \Omega_{\bm k} = \sum_{j,n} \Omega_j \sin ( \bm k \cdot \bm L_n^{(j)} ), \label{eq:shell}
\end{equation}
where $j$ runs over shells and $n=1,2,3$ runs over lattice vectors $\bm L_n^{(j)}$ related by $\mathcal C_{3z}$, see Fig.\ \ref{fig:fig2}(a). All shells are regular hexagons obtained by scaling and rotating the 1st shell. Since $\mathcal T$ is preserved, $\varepsilon_{\bm k}$ ($\Omega_{\bm k}$) is an even (odd) function of momentum. Crystal symmetries $\mathcal S$ act as $\varepsilon_{\bm k} = \varepsilon_{\mathcal S \bm k}$ and $\Omega_{\bm k} = \det(\mathcal S) \Omega_{\mathcal S \bm k}$ and thus constrain the coefficients $\varepsilon_j$ and $\Omega_j$. 

For example, the 2nd shell contains two lattice vectors related by $\mathcal C_{3z}$ and $\mathcal M_x$ (or $\mathcal C_{2y}$) symmetry which therefore contribute a term to $\Omega_{\bm k}$ that is even under $\mathcal M_x$. Hence $\Omega_2$ is forbidden by $\mathcal M_x$, but allowed by $\mathcal M_y$ (or $\mathcal C_{2x}$) which forbids $\Omega_1$. Similarly, only antisymmetric superpositions ($\Omega_4 = -\Omega_5$) of the degenerate 4th and 5th shells are allowed. Conversely, the dispersion conserves $\mathcal M_x$ for a symmetric superposition ($\varepsilon_4 = \varepsilon_5$). Plugging the expansions of \eqref{eq:shell} into \eqref{eq:JBloch} and \eqref{eq:Jgeom}, we obtain
\begin{align}
    J_\text{Bloch}^{\parallel / \perp} & = - \frac{6eL}{V_c \hbar} \sum_j \frac{\varepsilon_j f_j^0 L_j}{L} \, F_\text{Bloch}^{\parallel / \perp} \left( \omega_B \tau L_j / L , \theta + \theta_j \right), \\
    J_\text{geom} & = \frac{6eL}{V_c \tau} \sum_j \frac{\Omega_j f_j^0}{L^2} \frac{L}{L_j} \, F_\text{geom} \left( \omega_B \tau L_j / L , \theta + \theta_j \right),
\end{align}
where $f_j^0 = f_{\bm R}^0$ for $\bm R = \bm L_n^{(j)}$. Note that $\mathcal T$ implies $J_\text{Bloch}$ ($J_\text{geom}$) is even (odd) in $\bm E$. For the first five shells, the scaling factors $L_j/L = \{1, \sqrt{3}, 2, \sqrt{7}, \sqrt{7}\}$ and angles $\theta_j = \{0, \tfrac{\pi}{6}, 0, \tfrac{\pi}{6}+\varphi, \tfrac{\pi}{6}-\varphi\}$ with $\varphi = \arctan(\tfrac{\sqrt{3}}{5})$, see Fig.\ \ref{fig:fig2}(a). We also defined
\begin{align}
    F_\text{Bloch}^\parallel (\zeta,\theta) & = \frac{\zeta \left[ 8 + 6 \zeta^2 + \zeta^4 \cos^2(3\theta) \right]}{16 + 24 \zeta^2 + 9 \zeta^4 + \zeta^6 \cos^2(3\theta)}, \label{eq:JBpar} \\
    F_\text{Bloch}^\perp (\zeta,\theta) & = \frac{\zeta^5 \sin(3\theta) \cos(3\theta)}{16 + 24 \zeta^2 + 9 \zeta^4 + \zeta^6 \cos^2(3\theta)}, \label{eq:JBtr} \\
    F_\text{geom} (\zeta,\theta) & = \frac{\zeta^4 \left( 4 + \zeta^2 \right) \cos(3\theta)}{16 + 24 \zeta^2 + 9 \zeta^4 + \zeta^6 \cos^2(3\theta)}, \label{eq:Jg}
\end{align}
which are nonperturbative in the field strength. All material details are contained in the coefficients $\varepsilon_j$, $f_j^0$, and $\Omega_j$.
For a fixed field strength $E$, the currents are \emph{roses} as a function of the field direction $\theta$. The longitudinal rose, shown in Fig.\ \ref{fig:fig2}(b), only has one petal and depends rather weakly on $\theta$. On the other hand, the transverse roses are strongly anisotropic. Both $F_\text{Bloch}^\perp$ [Fig.\ \ref{fig:fig2}(c)] and $F_\text{geom}$ [Fig.\ \ref{fig:fig2}(d)] have six petals due to $\mathcal M_x$ symmetry, while $F_\text{Bloch}^\perp$ has an extra six petals from $\mathcal M_x \mathcal T$. As a check, we consider the weak-field limit ($\omega_B \tau \ll 1$) and recover the results from the symmetry analysis. As the field strength increases, $F_\text{Bloch}^\parallel$ attains a maximum at $(\omega_B \tau)^2 \in [4/3,2]$ depending on $\theta$, and decays as $E^{-1}$ for $\omega_B \tau \gg 1$. This decrease originates from electrons being Bragg reflected before relaxing to equilibrium and manifests as a negative differential conductance \cite{Esaki1970}. However, unlike the Bloch current, the geometric current plateaus for strong fields \cite{Phong2022b}. This is because the anomalous velocity grows linearly with the field, which cancels the $E^{-1}$ decay. Hence the geometric current dominates the response at strong field:
\begin{equation}
    \lim_{\omega_B \tau \gg 1} J_\text{geom} = \frac{6e}{V_c \tau} \sum_j \frac{\Omega_j f_j^0 \delta_{\theta+\theta_j,\pi/6+m\pi/3}}{L_j \cos[3(\theta + \theta_j)]},
\end{equation}
with $m \in \mathds Z$. The plateau value is very singular near $3(\theta+\theta_j) = \pi/2 + m\pi$ where contributions from $\Omega_j$ vanish. Terms with different $\theta_j$ can thus be distinguished by their dependence on the field direction. Moreover, because the geometric current first increases in magnitude and then plateaus, the geometric differential conductance attains an extremum. For the first shell, it lies at $\omega_B\tau \approx 1.477$ for $\theta=m\pi/3$ and shifts to larger fields when $\theta \rightarrow \pi/6 + m\pi/3$.

Experimentally, one can distinguish between the Bloch and geometric currents since they are odd and even in the electric field when $\mathcal T$ is conserved, respectively. Moreover, assuming the first shell dominates, which holds in the case of periodically-buckled graphene (see Material Systems), the ratios
\begin{align}
    \frac{J_\text{Bloch}^\perp}{J_\text{Bloch}^\parallel} & \simeq \frac{( \omega_B \tau )^4 \sin(3\theta) \cos(3\theta)}{8 + 6 ( \omega_B \tau )^2 + ( \omega_B \tau )^4 \cos^2(3\theta)}, \label{eq:ratio1} \\
    \frac{J_\text{Bloch}^\perp}{J_\text{geom}} & \simeq -\frac{\varepsilon_1 \tau L^2}{\hbar \Omega_1} \frac{\omega_B \tau \sin(3\theta)}{4 + ( \omega_B \tau )^2}, \label{eq:ratio2} 
\end{align}
are independent of the chemical potential and yield $\tau$ and $\varepsilon_1 / \Omega_1$. Measuring $J_\text{geom}$ at different fillings of the band would then, in principle, yield $\Omega_1$ and $\varepsilon_1$. The inverse problem of extracting the Fourier coefficients from the response is generally more tractable when $\Omega_{\bm k}$ and $\varepsilon_{\bm k}$ are sufficiently smooth such that only a limited number of shells contribute.

\section{Material Systems}

We apply the nonperturbative response theory to PBG ($C_{3v}$) and TDBG ($D_3$ and $C_3$). In both systems the band structure is tunable by applying an electric field normal to the $xy$ plane. We calculate $\varepsilon_\text{gap}^2 / \varepsilon_\text{width}$ for the highest valence band and the lowest conduction band of a given valley, and find broad windows in the strong-field regime where the band-projected theory is valid, i.e., 
\begin{equation}
    \frac{0.66 \, \text{ps}}{\tau}\frac{10 \, \text{nm}}{L} \ll \frac{E}{\text{kV/cm}} \ll \frac{\varepsilon_\text{gap}^2}{\varepsilon_\text{width} \text{meV}} \frac{10 \, \text{nm}}{L}.
\end{equation}
Other potential material realizations include periodically-gated Bernal bilayer graphene \cite{Ghorashi2022} and moir\'es based on transition-metal dichalcogenides \cite{Mak2022}. Moreover, large nonlinear responses have already been studied both theoretically and experimentally in twisted bilayer graphene, where a second-order Hall effect is possible when both $\mathcal C_{2z}$ and $\mathcal C_{3z}$ are broken either due to disorder \cite{Duan2022} or strain \cite{Pantaleon2021,Pantaleon2022,Zhang2022}.

We further consider the case where the buckling pattern (PBG) or the moir\'e lattice (TDBG) vary slowly with respect to the atomic lattice. Hence, the two valleys of graphene ($K_\pm$) are effectively decoupled. The total current is then obtained by summing contributions from both valleys, resulting in a small modification of the expressions in \eqref{eq:JBloch} and \eqref{eq:Jgeom}. Since the valleys are related by $\mathcal T$, the total current is obtained by letting $f_{\bm R}^0 \varepsilon_{-\bm R} \mapsto 2 \, \text{Re} \left( f_{\bm R}^{0+} \varepsilon_{-\bm R}^+ \right)$ in the Bloch current and $f_{\bm R}^0 \Omega_{-\bm R} \mapsto 2 \, \text{Im} \left( f_{\bm R}^{0+} \Omega_{-\bm R}^+ \right)$ in the geometric current where the superscript corresponds to valley $K_+$. This modification does not change  \eqref{eq:ratio1} but gives an extra factor from the phases of the Fourier components in \eqref{eq:ratio2}. A shell expansion for two decoupled bands, which individually break $\mathcal T$ and carry valley Chern numbers, is given in the SI (section II E). 

\subsection{Periodically-Buckled Graphene}

When monolayer graphene is placed on top of NbSe$_2$ or hBN \cite{Mao2020,Milovanovic2020}, as well as artificial nanobubble \cite{Qi2014} or nanopillar \cite{Jiang2017,Kang2021,Phong2022} substrates, it can undergo a buckling transition. Here we consider a substrate-induced buckling transition that gives rise to a periodic height profile with $C_{3v}$ symmetry. In the first-star approximation, the height profile is given by $h(\bm r) = h_0 \sum_{n=1}^3 \cos \left( \bm{\mathcal G}_n \cdot \bm r + \frac{\pi}{4} + \phi \right)$ with $\bm{\mathcal G}_n = \mathcal C_{3z}^{n-1} (0, 4\pi/\sqrt{3}L)$ where the phase $\phi$ controls the shape of the profile \cite{Mao2020,Phong2022,DeBeule2023,Mahmud2023}. Experimentally, $\phi$ can be tuned by designing different artificial substrates. When $h(\bm r)$ varies slowly on the graphene lattice scale ($L \gg a = 0.246 \; \text{nm}$) a valley-projected theory can be used with Hamiltonian \cite{Phong2022}
\begin{equation}
    \mathcal H_\nu = \hbar v_F \left[ \bm k + \frac{\nu e}{\hbar} \bm{\mathcal A}(\bm r) \right] \cdot \left( \nu \sigma_x, \sigma_y \right) + \mathcal V(\bm r) \sigma_0,
\end{equation}
where $\nu = \pm 1$ indicates the valleys $K_\nu$ and $v_F = \sqrt{3} \, t_0 a / 2$ is the Fermi velocity with $t_0 = 2.7$~eV \cite{Castro2009}. Here the scalar field $\mathcal V = \mathcal V_0 \sum_{n=1}^3 \cos \left( \bm{\mathcal G}_n \cdot \bm r + \frac{\pi}{4} + \phi \right)$ originates from applying an electric field (different from the driving field) normal to the nominal graphene plane \cite{Gao2023} and $\bm{\mathcal B} = \nabla \times \bm{\mathcal A} = \hat z \mathcal B(\bm r)$ with $\mathcal B(\bm r) = \mathcal B_0 \sum_{n=1}^3 \cos \left( \bm{\mathcal G}_n \cdot \bm r - 2\phi \right)$ the strain-induced pseudomagnetic field (PMF). The latter is obtained by taking into account in-plane relaxation while keeping the height modulation fixed \cite{Phong2022} (see SI, section III A). Up to a translation, the PMF is invariant under $\phi \mapsto \phi + \pi/3$ while $\mathcal V$ changes sign. Hence we restrict ourselves to $\phi \in (-\pi/6, \pi/6]$. For concreteness, we take $L/l_0 = 6$ with $L = 14 \; \text{nm}$ where $l_0 = \sqrt{\hbar / e\mathcal B_0} \propto \sqrt{aL^3/h_0^2}$ is an effective magnetic length. These are the experimental values of Ref.\ \cite{Mao2020}. Furthermore, because $\sigma_z \mathcal H_\nu[\mathcal V] \sigma_z = - \mathcal H_\nu[-\mathcal V]$, we only consider the highest valence band (of both valleys) for PBG.

\begin{figure}[t!]
    \centering
    \includegraphics[width=\linewidth]{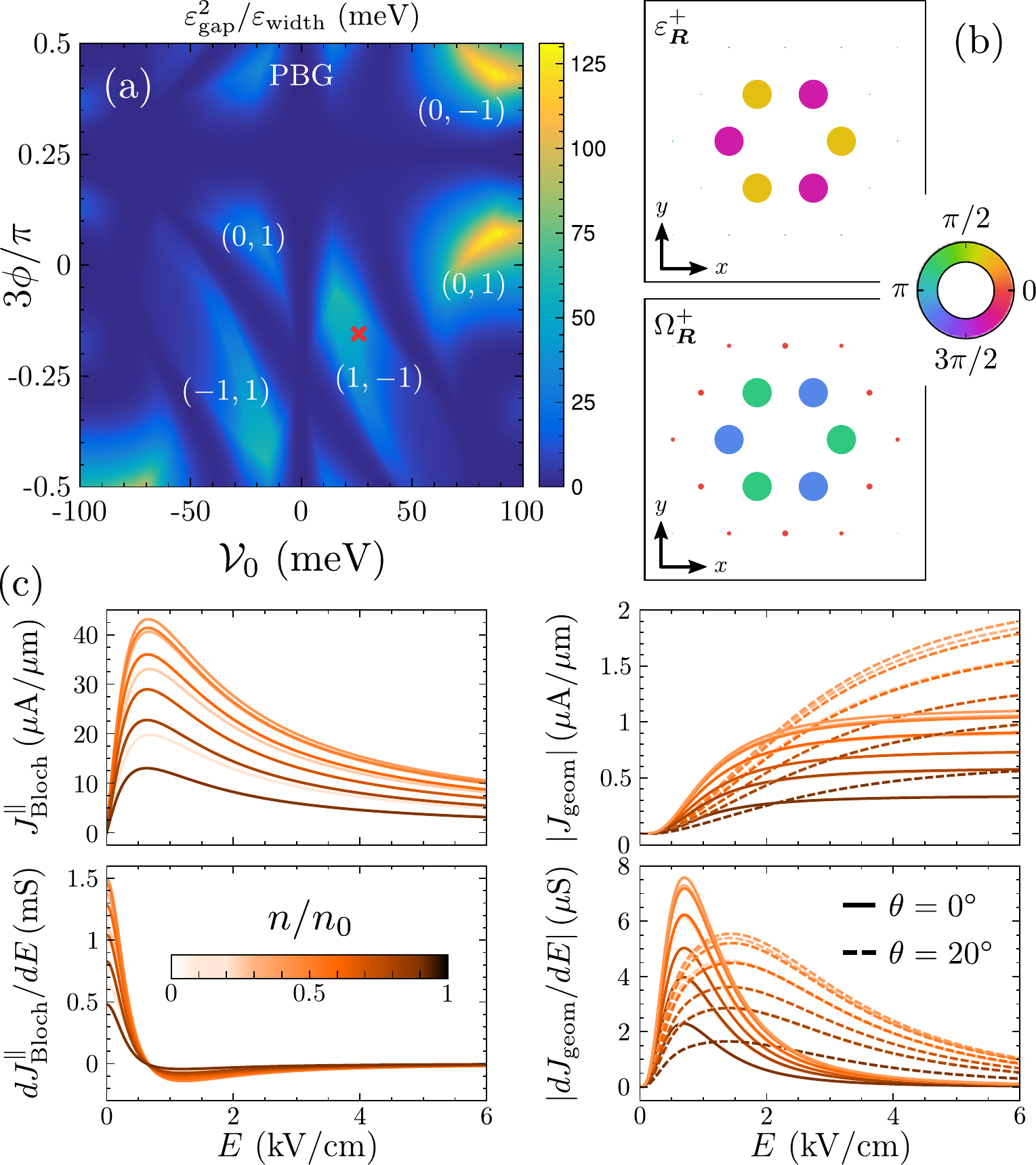}
    \caption{(a) $\varepsilon_\text{gap}^2/\varepsilon_\text{width}$ for the highest valence band of PBG in the $(\mathcal V_0,\phi)$ plane. The cross corresponds to the case shown in Fig.\ \ref{fig:fig1}(a). Some $K_+$ valley Chern numbers of the two bands near charge neutrality are shown. (b) Relative magnitude and phase of $\varepsilon_{\bm R}^+$ and $\Omega_{\bm R}^+$ for the cross in (a). (c) $J_\text{Bloch}^\parallel$ for field direction $\theta=0^\circ$ and $|J_\text{geom}|$ for $\theta=0^\circ$ (solid) and $\theta=20^\circ$ (dashed), as well as the differential conductance, for $T = 5$~K. The color scale gives the filling $n/n_0$ of the band.}
    \label{fig:fig3}
\end{figure}

In Fig.\ \ref{fig:fig3}(a), we show $\varepsilon_\text{gap}^2/\varepsilon_\text{width}$ in the $(\mathcal V_0, \phi)$ plane for the highest valence band. By varying the shape of the height profile and the electric field normal to the nominal graphene plane, this ratio is in the range $10-100 \; \text{meV}$ which should be large enough to avoid electric breakdown in the strong-field regime. As an example, we consider the parameters indicated with a cross on Fig.\ \ref{fig:fig3}(a). For this case, the valley Chern number is given by $\pm 1$ for valley $K_\pm$ and the bands along high-symmetry lines are shown in Fig.\ \ref{fig:fig1}(a). In Fig.\ \ref{fig:fig3}(b), we show the relative magnitude and phase of the Fourier components $\varepsilon_{\bm R}^+$ and $\Omega_{\bm R}^+$. We see that the first-shell dominates in this particular case. The longitudinal and geometric current, as well as the corresponding differential conductance are shown in Fig.\ \ref{fig:fig3}(c) for different fillings of the band. We find that the strong-field regime is reached for $E \approx 1 \; \text{kV}/\text{cm}$. The longitudinal current does not depend strongly on the field direction, hence we only show the case $\theta = 0^\circ$. Here the $E^{-1}$ decay for $\omega_B \tau \gg 1$ manifests as a negative differential conductance. On the other hand, the geometric current is strongly anisotropic, as is clear from the current rose shown in Fig.\ \ref{fig:fig1}(c). We see that the plateau in the geometric current shifts to larger fields as we increase $\theta$ from $0^\circ$ to $20^\circ$, concomitant with a shift and broadening of the peak in the differential conductance.

\subsection{Twisted Double Bilayer Graphene}

TDBG consists of a stack of two (AB or BA) Bernal bilayer graphene layers that are twisted relative to each other \cite{Chebrolu2019,Koshino2019}. A second-order Hall effect was recently observed in TDBG for twist angles $\theta \sim 1^\circ$ where $\mathcal C_{3z}$ was broken by strain \cite{Sinha2022,Chakraborty2022,zhong2023effective,Tiwari2023}, making this system a promising platform for studying nonperturbative responses. For such small twists one can again use a valley-projected theory, see SI (section IV). The tunable parameters for TDBG are now given by the twist angle $\vartheta$ and the bias difference $U$ between the topmost and bottommost layer due to an applied electric field (different from the driving field) normal to the TDBG plane. The latter reduces the point group of TDBG from $D_3$ to $C_3$. As such, the current roses are less constrained and only show $\mathcal C_{3z}$ symmetry, see Fig.\ \ref{fig:fig1}(d). 

\begin{figure}[t!]
    \centering
    \includegraphics[width=\linewidth]{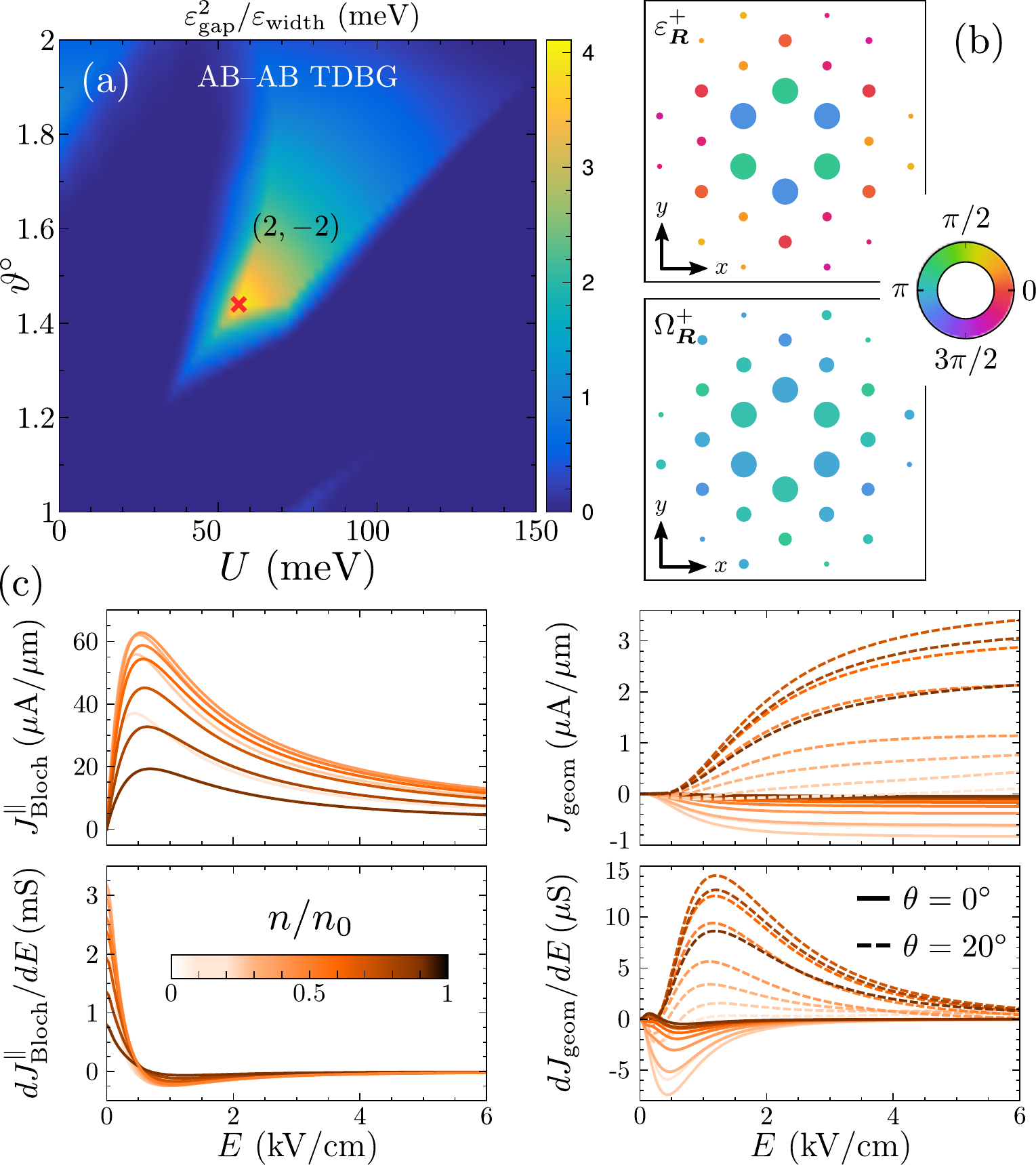}
    \caption{(a) $\varepsilon_\text{gap}^2/\varepsilon_\text{width}$ for the lowest conduction band of AB--AB TDBG in the $(U,\vartheta)$ plane. The cross gives the case shown in Fig.\ \ref{fig:fig1}(b) and the $K_+$ valley Chern numbers of the two bands near charge neutrality are shown. (b) Relative magnitude and phase of $\varepsilon_{\bm R}^+$ and $\Omega_{\bm R}^+$ for the cross in (a) up to the fifth shell. (c) $J_\text{Bloch}^\parallel$ for field direction $\theta=0^\circ$ and $J_\text{geom}$ for $\theta=0^\circ$ (solid) and $\theta=20^\circ$ (dashed), as well as $dJ/dE$, for $T = 5$~K. The color scale gives the filling $n/n_0$ of the band.}
    \label{fig:fig4}
\end{figure}

In Fig.\ \ref{fig:fig4}(a), we show $\varepsilon_\text{gap}^2/\varepsilon_\text{width}$ in the $(U, \vartheta)$ plane for the lowest conduction band. By varying the twist angle and the bias, this ratio can be of the order of $5 \; \text{meV}$ which limits the range of electric fields where the band-projected theory is valid in the strong-field regime to a few $\text{kV/cm}$. We note that this ratio can be larger for smaller twist angles $\vartheta \ll 1^\circ$. However, for such small twists, lattice relaxation might become important and as such we do not consider them here. As an example, we consider the parameters indicated with a cross on Fig.\ \ref{fig:fig4}(a). For this case, the valley Chern number is given by $\mp 2$ for valley $K_\pm$ and the bands along high-symmetry lines are shown in Fig.\ \ref{fig:fig1}(b). In Fig.\ \ref{fig:fig4}(b), we show the relative magnitude and phase of the Fourier components $\varepsilon_{\bm R}^+$ and $\Omega_{\bm R}^+$ up to the fifth shell. Contrary to the case chosen for PBG in Fig.\ \ref{fig:fig3}, many shells contribute. The longitudinal and geometric current, as well as the corresponding differential conductance are shown in Fig.\ \ref{fig:fig4}(c) for different fillings of the band. Because the moir\'e lattice constant $L_m(\vartheta=1.44^\circ) \approx 9.8 \; \text{nm}$ is of the same order as the one chosen for PGB and the first shell is still the largest contribution, the onset of the strong-field regime is again given by $E \approx 1 \; \text{kV}/\text{cm}$. Generically, the longitudinal current does not depend strongly on the field direction, hence we only show the case $\theta = 0^\circ$. On the other hand, as already demonstrated in Fig.\ \ref{fig:fig1}(d), the geometric current is strongly anisotropic. The field strength for which $J_\text{geom}$ plateaus, as well as the position and width of the peak in $dJ_\text{geom}/dE$ is strongly dependent on the field direction. Moreover, since the point group of interlayer-biased TDBG is $C_3$ there are no mirror axes for which the transverse currents vanish. Hence there is no fixed field direction for which the plateau is reached first as a function of the field strength.

\section{Discussion}

In this work, we have studied the anistropy in the current response to a static electric field within a semiclassical band-projected theory up to infinite order in the field strength. We have focused on two-dimensional superlattice systems with trigonal symmetry that feature spectrally isolated and narrow minibands, for which electric breakdown is absent even in the strong-field regime. We have demonstrated that the Bloch (geometric) currents can be written in terms of an expansion in coordination shells where each term is given by a rose weighted by the Fourier component of the group velocity (Berry curvature). Here, each rose obeys the symmetries of the given shell. While the longitudinal current rose has no petals and hence a weak anisotropy, the transverse currents are strongly anisotropic. The latter follows from the fact that roses originating from shells with mirror symmetries necessarily have nodes and thus several petals. Furthermore, for the longitudinal and Bloch transverse response, the strong-field regime is characterized by a negative differential conductance due to electrons being Bragg reflected before relaxing their momentum. For the geometric response, however, the current plateaus in the strong-field regime, yielding a peak in the differential conductance whose position and width strongly depend on the field direction. 

We have suggested two candidate systems: periodically-buckled graphene and twisted double bilayer graphene. For these systems, strong-field responses are accessible at field strengths $E \sim 1 \text{ kV/cm}$ owing to a lattice constant of the order of $10 \; \text{nm}$. Importantly, because these systems break $\mathcal C_{2z}$ symmetry but conserve time-reversal symmetry, there is a nonlinear geometric response from the momentum distribution of Berry curvature. We have further shown that $\varepsilon_\text{gap}^2/\varepsilon_\text{width} \sim 50 \; \text{meV}$ for PBG and $\varepsilon_\text{gap}^2/\varepsilon_\text{width} \sim 5 \; \text{meV}$ for AB--AB TDBG, such that especially for the case of PBG, the strong-field regime can be reached well before electric breakdown. Finally, we note that most moir\'e systems display some degree of heterostrain which breaks rotation symmetry \cite{Xie2019}. In some cases, this  feature is necessary to observe a second order Hall effect \cite{Sinha2022,Chakraborty2022,zhong2023effective,Tiwari2023}. Hence, in the presence of strain, the symmetry of the rose pattern at low fields is expected to be reduced, while at larger fields, the petal structure enforced by $\mathcal C_{3z}$ symmetry is expected to be only slightly perturbed.

\begin{acknowledgements}
We thank S.\ Gassner for discussions. This research was funded in whole, or in part, by the Luxembourg National Research Fund (FNR project No.\ 16515716). Work by CDB, VTP, and EJM is supported by the Department of Energy under grant DE-FG02-84ER45118. VTP acknowledges further support from the P.D. Soros Fellowship for New Americans and the National Science Foundation's Graduate Research Fellowships Program.
\end{acknowledgements}

\bibliography{references}
\let\addcontentsline\oldaddcontentsline 

\onecolumngrid
\newpage
\begin{center}
\textbf{\large Supplementary Information for ``Roses in the Nonperturbative Current Response of Artificial Crystals''}
\end{center}
\setcounter{equation}{0}
\setcounter{figure}{0}
\setcounter{table}{0}
\setcounter{page}{1}
\setcounter{section}{0}
\makeatletter
\renewcommand{\thepage}{S\arabic{page}}
\renewcommand{\thesection}{S\arabic{section}}
\renewcommand{\theequation}{S\arabic{equation}}
\renewcommand{\thefigure}{S\arabic{figure}}

\tableofcontents

\section{Current response to a uniform electric field}

We give an overview of the calculation of the steady-state current in a uniform static electric field. We start from the semiclassical equations of motion and the Boltzmann transport equation in the band-projected theory. We then proceed to evaluate this expression by expanding the band dispersion and the Berry curvature in coordination shells.

\subsection{Semiclassical electron dynamics}

The semiclassical equations of motion for an electron in a two-dimensional (2D) crystal, occupying an energy band with dispersion $\varepsilon_{n\bm k}$ subjected to a uniform and static electric field $\bm E$ are given by \cite{Chang1995,Sundaram1999}
\begin{align}
    \hbar \dot{\bm r}_{n\bm k} & = \nabla_{\bm k} \varepsilon_{n\bm k} - \hbar \dot{\bm k} \times \bm \Omega_{n\bm k}, \\
    \hbar \dot{\bm k} & = -e \bm E,
\end{align}
with $n$ the band index and $\bm \Omega_{n\bm k} = \Omega_{n\bm k} \hat z$ the Berry curvature, defined as
\begin{equation}
    \Omega_{n\bm k} = i \left( \left\langle \frac{\partial u_{n\bm k}}{\partial k_x} \bigg\rvert \frac{\partial u_{n\bm k}}{\partial k_y} \right\rangle_\text{cell} - \left\langle \frac{\partial u_{n\bm k}}{\partial k_y} \bigg\rvert \frac{\partial u_{n\bm k}}{\partial k_x} \right\rangle_\text{cell} \right),
\end{equation}
where $u_{n\bm k}(\bm r)$ are cell-periodic Bloch functions in periodic gauge, $u_{n,\bm k+\bm G}(\bm r) = e^{-i \bm G \cdot \bm r} u_{n\bm k}(\bm r)$ with $\bm G$ a reciprocal lattice vector, and $\langle u_{n\bm k} | u_{m\bm k} \rangle_\text{cell} = \delta_{nm}$. Here we have assumed that terms originating from interband transitions such as the field correction to the Berry curvature \cite{Gao2014} can be neglected. In the following, we omit the band index $n$ since we consider a single band. 

The current density is given by 
\begin{equation}
    \bm J = -e \int_\text{BZ} \frac{d^2\bm k}{\left( 2 \pi \right)^2} \, f_{\bm k} \dot{\bm r}_{\bm k} \equiv \bm J_\text{Bloch} + \bm J_\text{geom},
\end{equation}
with $e>0$ the elementary charge and \cite{Phong2022b}
\begin{align}
    \bm J_\text{Bloch} & = -\frac{e}{\hbar} \int_\text{BZ} \frac{d^2\bm k}{\left( 2 \pi \right)^2} \, f_{\bm k} \nabla_{\bm k} \varepsilon_{\bm k}, \\
    \bm J_\text{geom} & = \left( \hat z \times \bm E \right) \frac{e^2}{\hbar} \int_\text{BZ} \frac{d^2\bm k}{\left( 2 \pi \right)^2} \, f_{\bm k} \Omega_{\bm k},
\end{align}
where $f_{\bm k}$ is the out-of-equilibrium distribution function, obtained from the Boltzmann equation. Note that unlike in the main text, we do \emph{not} add the factor of $2$ for spin in the Supplementary Information. In the relaxation-time approximation, the Boltzmann equation is given by
\begin{equation} \label{eq:boltzmann}
    \frac{\partial f}{\partial t} + \dot{\bm k} \cdot \frac{\partial f}{\partial \bm k} + \dot{\bm r} \cdot \frac{\partial f}{\partial \bm r} = - \frac{f - f^0}{\tau},
\end{equation}
where $\tau$ is the momentum-relaxation time and $f_{\bm k}^0 = f^0(\varepsilon_{\bm k})$ is the Fermi function, 
\begin{equation}
    f^0(\varepsilon) = \frac{1}{e^{\left( \varepsilon - \mu \right) / k_B T} + 1},
\end{equation}
with $\mu$ the chemical potential and $T$ the temperature. We are interested in the steady-state response of a uniform electric field, such that the first and third term on the left-hand side of Eq.\ \eqref{eq:boltzmann} vanish. Hence, we obtain
\begin{equation}
    f_{\bm k} - \frac{e \tau}{\hbar} \bm E \cdot \frac{\partial f_{\bm k}}{\partial \bm k} = f_{\bm k}^0,
\end{equation}
which is formally solved by
\begin{equation}
    f_{\bm k} = f_{\bm k}^0 + \frac{e \tau}{\hbar} E_i \frac{\partial f_{\bm k}^0}{\partial k_i} + \left( \frac{e \tau}{\hbar} \right)^2 E_i E_j \frac{\partial^2 f_{\bm k}^0}{\partial k_i k_j} + \cdots.
\end{equation}
For a translational-invariant system, $f_{\bm k}^0$ can be expanded as a Fourier series,
\begin{equation}
    f_{\bm k}^0 = \sum_{\bm R} f_{\bm R}^0 \, e^{i \bm k \cdot \bm R}, \qquad \qquad f_{\bm R}^0 = \frac{V_c}{\left( 2 \pi \right)^2} \int_\text{BZ} d^2\bm k \, f_{\bm k}^0 \, e^{-i \bm k \cdot \bm R},
\end{equation}
where $\bm R$ are lattice vectors and $V_c$ the area of the unit cell. This yields
\begin{equation}
    f_{\bm k} = \sum_{\bm R} \frac{f_{\bm R}^0 \, e^{i \bm k \cdot \bm R}}{1 - i e \tau \bm E \cdot \bm R / \hbar}.
\end{equation}
Plugging this result for the distribution function back into the expression for the currents, we obtain
\begin{align}
    \bm J_\text{Bloch}(\bm E) & = -\frac{e}{\hbar} \sum_{\bm R, \bm R'} \frac{i \bm R' f_{\bm R}^0 \varepsilon_{\bm R'}}{1 - i e \tau \bm E \cdot \bm R / \hbar} \, \int_\text{BZ} \frac{d^2\bm k}{\left( 2 \pi \right)^2} \, e^{i \bm k \cdot ( \bm R + \bm R' )}, \\
    & = \frac{e}{V_c \hbar} \sum_{\bm R} \frac{i \bm R f_{\bm R}^0 \varepsilon_{-\bm R}}{1 - i e \tau \bm E \cdot \bm R / \hbar}, \\
    \bm J_\text{geom}(\bm E) & = \left( \hat z \times \bm E \right) \frac{e^2}{V_c \hbar} \sum_{\bm R} \frac{f_{\bm R}^0 \Omega_{-\bm R}}{1 - i e \tau \bm E \cdot \bm R / \hbar}.
\end{align}

\subsection{Symmetry properties of the current}

We now discuss the constraints put on the currents by symmetry. We start with time-reversal ($\mathcal T$) symmetry. In the presence of $\mathcal T$, the band dispersion $\varepsilon_{\bm k}$ is an even function of momentum, while the Berry curvature $\Omega_{\bm k}$ is odd. Hence in real space, we have $\varepsilon_{\bm R} = \varepsilon_{-\bm R}$ and $f_{\bm R}^0 = f_{-\bm R}^0$,  while $\Omega_{\bm R} = -\Omega_{-\bm R}$. We thus see that in a time-reversal-invariant system, $\varepsilon_{\bm R}$ and $f_{\bm R}^0$ are real while $\Omega_{\bm R}$ is imaginary. This then implies
\begin{align}
    \bm J_\text{Bloch} (\bm E) & \overset{\mathcal T}{=} - \bm J_\text{Bloch}(-\bm E), \\
    \bm J_\text{geom} (\bm E) & \overset{\mathcal T}{=} \bm J_\text{geom}(-\bm E).
\end{align}
Hence, when $\mathcal T$ is preserved, the geometric (Bloch) current gives that part of the current that is even (odd) in the electric field.

If the system conserves a crystalline symmetry $\mathcal S$, the current obeys
\begin{equation} \label{eq:Jtransform}
    \bm J( \mathcal S \bm E) \overset{\mathcal S}{=} \mathcal S \bm J(\bm E).
\end{equation}
For example, we see that when $\mathcal C_{2z}$ [$(x,y) \mapsto (-x,-y)$] is conserved, the total current is odd in the electric field. In combination with time-reversal symmetry, this implies that the geometric current vanishes, consistent with the fact that the Berry curvature vanishes in that case. Likewise, under a mirror symmetry $\mathcal M_x$ ($x \mapsto -x$),
\begin{align}
    J_x (E_x, E_y) & = - J_x(-E_x,E_y), \\
    J_y (E_x, E_y) & = + J_y(-E_x,E_y),
\end{align}
such that $J_x$ vanishes for $E_x=0$. Hence a transverse response is forbidden whenever the electric field lies along a mirror axis. In general, the longitudinal and transverse components of the current transform as
\begin{align}
    J_\parallel(\bm E) & \equiv \hat E \cdot \bm J(\bm E) = \mathcal S \hat E \cdot \bm J(\mathcal S \bm E) = J_\parallel(\mathcal S \bm E) \\
    J_\perp(\bm E) & \equiv ( \hat E \times \hat z ) \cdot \bm J(\bm E) = \det(\mathcal S) J_\perp(\mathcal S \bm E),
\end{align}
where $\bm E = E \hat E$ and $\bm J = J_\parallel \hat E + J_\perp \hat E \times \hat z$. Here we used Eq.\ \eqref{eq:Jtransform}. Hence, the longitudinal component transforms as a scalar field, while the transverse component transforms as a pseudoscalar field. Note that out-of-plane rotations, such as $\mathcal C_{2x}$, act as improper rotations when restricted to the $xy$ plane with $\det \mathcal S = -1$.

\subsection{Weak-field expansion} 

Here we obtain a series expansion in powers of the electric field for the longitudinal and transverse components from symmetry considerations. To this end, we first define the even and odd currents,
\begin{equation}
    \bm J^{(\pm)} (\bm E) = \frac{\bm J(\bm E) \pm \bm J(-\bm E)}{2}.
\end{equation}
In the presence of a rotation symmetry about the principal axis $\mathcal C_{nz}$ ($n=2,3,4,6$) of the 2D crystal, we see that the even component is present only for $n=3$. 

In order to implement the symmetry, we need and object that transforms properly under the symmetry. For the rotation symmetry, we consider the object $J_x + iJ_y = Je^{i\theta}$ with $\hat E = (\cos\theta,\sin\theta)$. Hence this object transforms as an $L_z = 1$ object under $\mathcal C_{nz}$. We first consider the odd component and focus on $\mathcal C_{3z}$ symmetry. Up to fifth order in the electric field as, we can write
\begin{equation}
    J_x^{(-)} + i J_y^{(-)} = a(E^2) \left( E_x + i E_y \right) + b \left( E_x - i E_y \right)^5 + \mathcal O(E^7),
\end{equation}
where both sides transform as an $L_z = 1$ object. Here we used that $L_z$ is only conserved mod $3$ for a system with $\mathcal C_{3z}$ symmetry. Incidentally, for a system with $\mathcal C_{6z}$, there are no extra terms and we obtain the same expression for the odd current. Here we defined the functions
\begin{equation}
    a = a_0 + a_1 E^2 + a_2 E^4, \qquad b = b_0,
\end{equation}
with $a_0$, $a_1$, $a_2$, and $b_0$ c-numbers. We note that $a_0 = \sigma_L + i \sigma_H$ where $\sigma_L$ ($\sigma_H$) is the linear longitudinal (Hall) conductivity. For a system with $\mathcal T$ symmetry, $a$ is real because of Onsager reciprocity. Moreover, a mirror or in-plane rotation axis in the $y$ direction, imply that the functions $a$ and $b$ are real. Projecting in the directions parallel $\hat E = \left( \cos \theta, \sin \theta \right)$ and perpendicular $\hat z \times \hat E = \left( -\sin \theta, \cos \theta \right)$ the electric field, yields
\begin{equation}
    J_\parallel^{(-)} - i J_\perp^{(-)} \simeq a E + b E^5e^{-i6\theta},
\end{equation}
which, without taking into account any symmetry other than $\mathcal C_{3z}$, gives
\begin{align}
    J_\parallel^{(-)} & \simeq \text{Re}(a_0) E + \text{Re}(a_1) E^3 + \left[ \text{Re}(a_2) + | b_0 | \cos(6\theta - \arg b_0) \right] E^5, \\
    -J_\perp^{(-)} & \simeq \text{Im}(a_0) E + \text{Im}(a_1) E^3 + \left[ \text{Im}(a_2) - |b_0| \sin(6\theta - \arg b_0) \right] E^5.
\end{align}
Note that the \emph{projected} even (odd) current is actually odd (even) in the electric field. We thus find that the anisotropy in the odd current only emerges at fifth order in the electric field. When the system has a mirror axis ($\mathcal M_x : x \mapsto -x$) or a rotation symmetry ($\mathcal C_{2y}$) about the $y$ axis, this reduces to
\begin{align}
    J_\parallel^{(-)} & \simeq a_0 E + a_1 E^3 + \left[ a_2 + b_0 \cos(6\theta) \right] E^5, \\
    J_\perp^{(-)} & \simeq b_0 E^5 \sin(6\theta),
\end{align}
where all coefficients are real. One can perform the same analysis for a system with $\mathcal C_{2z}$ or $\mathcal C_{4z}$ symmetry. For the former, we find
\begin{equation}
    J_\parallel^{(-)} - i J_\perp^{(-)} = a E + b E e^{-i2\theta} + \mathcal O(E^3),
\end{equation}
and for the latter,
\begin{equation}
    J_\parallel^{(-)} - i J_\perp^{(-)} = a E + b E^3 e^{-i4\theta} + \mathcal O(E^5).
\end{equation}
Here we expanded up to the lowest order that shows anisotropy.
\begin{table}
\renewcommand{\arraystretch}{1.5}
\setlength{\tabcolsep}{5pt}
\caption{Expansions of the currents that are even ($J^{(+)}$) and odd ($J^{(-)}$) in the electric field, up to leading order in the anisotropy, in the presence of $\mathcal C_{nz}$ symmetry ($n=2,3,4,6$). There are additional constraints on the functions $a(E^2)$ and $b$ if $\mathcal T$ or other crystalline symmetries are conserved.}
\begin{center}
\begin{tabular}{l | c | c}
\Xhline{1pt}
 & $J_x^{(+)} + i J_y^{(+)}$ & $J_x^{(-)} + i J_y^{(-)}$ \\
\Xhline{1pt}
$\mathcal C_{2z}$ only & $0$ & $a \left( E_x + i E_y \right) + b \left( E_x - i E_y \right)$ \\
\hline
$\mathcal C_{3z}$ & $a(E^2) \left( E_x - i E_y \right)^2 + b \left( E_x + i E_y \right)^4$ & $a(E^2) \left( E_x + i E_y \right) + b \left( E_x - i E_y \right)^5$ \\
\hline
$\mathcal C_{4z}$ & $0$ & $a(E^2) \left( E_x + i E_y \right) + b \left( E_x - i E_y \right)^3$ \\
\hline
$\mathcal C_{6z}$ & $0$ & $a(E^2) \left( E_x + i E_y \right) + b \left( E_x - i E_y \right)^5$ \\
\Xhline{1pt}
$\mathcal M_x$ or $\mathcal C_{2y}$ & $ia, ib \in \mathds R$ & $a, b \in \mathds R$ \\
\hline
$\mathcal M_y$ or $\mathcal C_{2x}$ & $a,b \in \mathds R$ & $a, b \in \mathds R$ \\
\hline
$\mathcal T$ & $a = -b^* E^2$ & $a \in \mathds R$ \\
\Xhline{1pt}
\end{tabular}
\label{tab:SIweak}
\end{center}
\end{table}

Similarly, we expand the even part of current up to fourth order in the electric field. Since the even part of the current vanishes in the presence of $\mathcal C_{2z}$ symmetry, we only need to consider $\mathcal C_{3z}$ symmetry:
\begin{equation}
    J_x^{(+)} + i J_y^{(+)} = a(E^2) \left( E_x - i E_y \right)^2 + b \left( E_x + i E_y \right)^4 + \mathcal O(E^6),
\end{equation}
with $a = a_0 + a_1 E^2$ and $b = b_0$. The longitudinal and transverse components become
\begin{equation}
    J_\parallel^{(+)} - i J_\perp^{(+)} \simeq a E^2 e^{-i3\theta} + b_0 E^4 e^{i3\theta}.
\end{equation}
In the semiclassical theory, the presence of $\mathcal T$ symmetry requires that $J_\parallel^{(+)}$ vanishes since the geometric current is purely transversal. This implies that $a_0 = 0$ and $a_1 = -b_0^*$. Moreover, a mirror axis along the $y$ direction further constrains $a$ and $b$ to be purely imaginary. Hence $a_1 = b_0 = -ic$ such that
\begin{equation}
    J_\perp^{(+)} \simeq 2 c E^4 \cos(3\theta).
\end{equation}
An overview of the weak-field expansions of the even and odd current is given in Table \ref{tab:SIweak}.

\section{Expansion in coordination shells}

In this section, we calculate the current with the semiclassical theory for a single isolated Chern trivial band. This band is part of a larger band manifold but is well-separated from other bands. We further assume that time-reversal ($\mathcal T$) symmetry, $\mathcal C_{3z}$ rotation symmetry, and  $\mathcal M_x$ ($x \mapsto -x$) mirror symmetry are preserved, but that $\mathcal C_{2z}$ or spatial inversion symmetry is broken. Hence, a finite Berry curvature is allowed even though the Chern number vanishes. 

We write the dispersion $\varepsilon_{\bm k}$ and the Berry curvature $\Omega_{\bm k}$ in terms of an expansion in the coordination shells of the triangular lattice:
\begin{align}
    \varepsilon_{\bm k} & = \varepsilon_0 + \varepsilon_1 \sum_{n=1}^3 \cos ( \bm k \cdot \bm L_n^{(1)} ) + \varepsilon_2 \sum_{n=1}^3 \cos ( \bm k \cdot \bm L_n^{(2)} ) + \cdots, \\
    \Omega_{\bm k} & = \Omega_1 \sum_{n=1}^3 \sin ( \bm k \cdot \bm L_n^{(1)} ) + \Omega_3 \sum_{n=1}^3 \sin ( \bm k \cdot \bm L_n^{(3)} ) + \cdots,\label{eq:Omega_k}
\end{align}
where we set $\varepsilon_0=0$ from now on. The first coordination shell is given by a regular hexagon whose vertices lie at a distance $L$ from the origin, where $L$ is the lattice constant. Here we choose 
\begin{align}
    \bm L_1^{(1)} & = \bm L_1, \\
    \bm L_2^{(1)} & = \bm L_2, \\
    \bm L_3^{(1)} & = - \left( \bm L_1 + \bm L_2 \right),
\end{align}
where $\bm L_1 = L \left( 1/2, \sqrt{3}/2 \right)$ and $\bm L_2 = L \left( -1, 0 \right)$ are primitive lattice vectors. Here we define the lattice vectors such that $\mathcal C_{3z}$, i.e., $\bm L_{n+1}^{(j)} = \mathcal C_{3z} \bm L_{n}^{(j)}$ for $j=1,2,3,\ldots$. The second shell is given by a regular hexagon that is rotated by $\pi/6$ with respect to the first shell and scaled by a factor of $\sqrt{3}$, as shown in Fig.\ 2(a) of the main text. The third shell is given by the first shell scaled by a factor of $2$. The corresponding lattice vectors can be chosen as
\begin{align}
    \bm L_1^{(2)} & = \bm L_1 - \bm L_2, \\
    \bm L_2^{(2)} & = \bm L_1 + 2 \bm L_2, \\
    \bm L_3^{(2)} & = - \left( 2 \bm L_1 + \bm L_2 \right), \\
    \bm L_1^{(3)} & = 2 \bm L_1, \\
    \bm L_2^{(3)} & = 2 \bm L_2, \\
    \bm L_3^{(3)} & = -2 \left( \bm L_1 + \bm L_2 \right).
\end{align}
The fourth and fifth shells are degenerate, i.e., they both lie at a distance $\sqrt{7} \, L$ from the origin. These shells are given by two regular hexagons that are rotated by an angle $\pi/6 \pm \arctan \left( \sqrt{3}/5 \right)$ with respect to the first shell, respectively. 

\subsection{Symmetry constraints}

Time-reversal symmetry requires an even band dispersion $\varepsilon_{\bm k} = \varepsilon_{-\bm k}$, while the Berry curvature is required to be odd, $\Omega_{\bm k} = -\Omega_{-\bm k}$. Under a crystalline symmetry $\mathcal S$, the band dispersion transforms as a scalar, while the Berry curvature transforms as a pseudoscalar:
\begin{equation}
    \varepsilon_{\bm k} = \varepsilon_{\mathcal S \bm k}, \qquad \Omega_{\bm k} = \det(\mathcal S) \Omega_{\mathcal S \bm k}.
\end{equation}
Hence, the Berry curvature acquires a sign under $\mathcal M_x$ symmetry. This constrains the coefficients $\varepsilon_j$ and $\Omega_j$ in the expansion in coordination shells, given in Eq.\ \eqref{eq:Omega_k}. For instance, the second shell in $\Omega_{\bm k}$ does not transform properly under $\mathcal M_x$, instead it transforms properly under $\mathcal M_y$. Hence we set $\Omega_2 = 0$ in Eq.\ \eqref{eq:Omega_k}. This can be understood from the shell structure shown in Fig.\ 2(a) of the main text. For the 2nd shell, there are two lattice vectors that are related by both $\mathcal C_{3z}$ and $\mathcal M_x$ symmetry, such that the corresponding term of the Berry curvature will be even under $\mathcal M_x$. The 3rd shell, however, is merely a rescaled version of the 1st shell and $\Omega_3$ can be therefore be finite. Similarly, only antisymmetric superpositions of the fourth and fifth shells are allowed, i.e., $\Omega_4 = -\Omega_5$. Conversely, the band dispersion requires a symmetric superposition of the 4th and 5th shells to conserve $\mathcal M_x$ symmetry, i.e., $\varepsilon_4 = \varepsilon_5$. Hence, we need to expand up to the 4th and 5th shells to break $\mathcal M_x$ in the band dispersion, while we only need to expand up to the 2nd shell to break $\mathcal M_x$ in the Berry curvature. Similar relations hold for higher-order shells.

\begin{figure}
    \centering
    \includegraphics[width=0.75\linewidth]{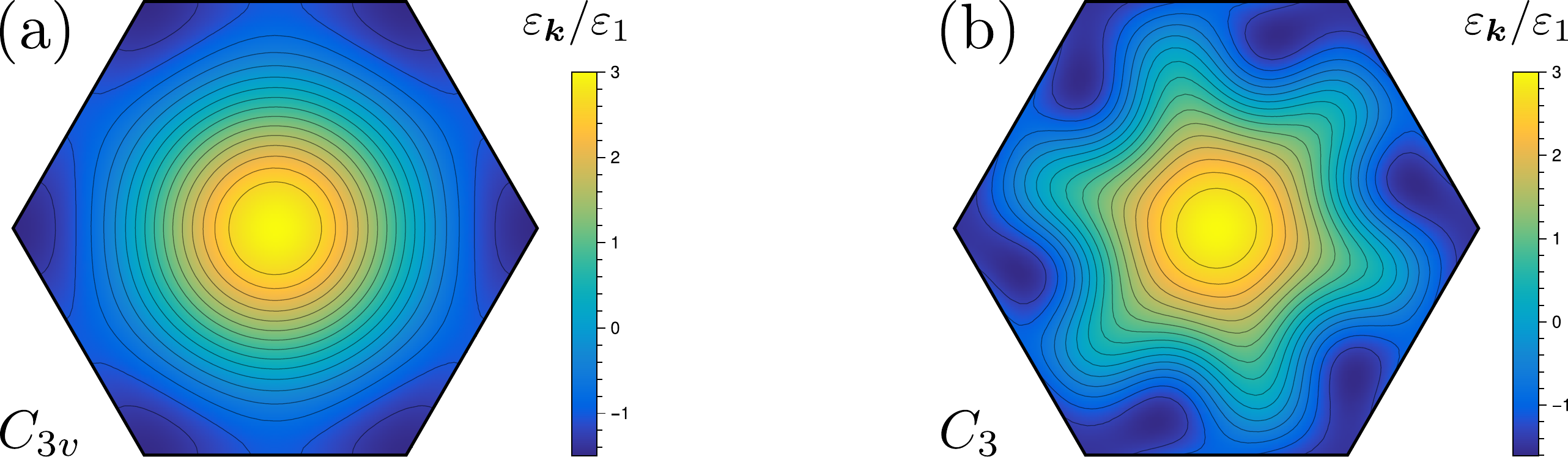}
    \caption{Energy bands with $\mathcal C_{3z}$ and $\mathcal T$ symmetry. (a) First-shell approximation, i.e., only $\varepsilon_1$ is finite such that $\mathcal M_x$ is conserved. (b) Including higher shells that explicitly break $\mathcal M_x$ symmetry, $\varepsilon_4/\varepsilon_1 = -\varepsilon_5/\varepsilon_1 = 0.2$.}
    \label{fig:SI_bands}
\end{figure}

\begin{figure}
    \centering
    \includegraphics[width=0.75\linewidth]{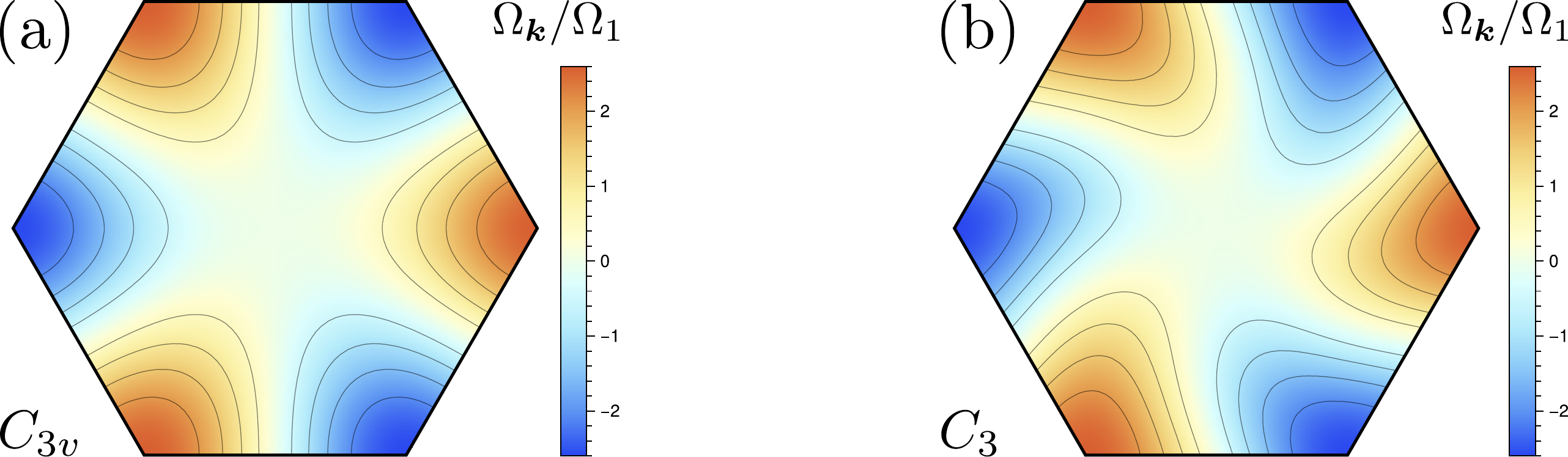}
    \caption{Berry curvature for $\mathcal C_{3z}$ and $\mathcal T$ symmetry, but broken $\mathcal C_{2z}$ or inversion symmetry. (a) First-shell approximation. Only $\Omega_1$ contributes and $\mathcal M_x$ is conserved. (b) Including the second shell breaks $\mathcal M_x$ symmetry, $\Omega_2/\Omega_1 = 0.2$.}
    \label{fig:SI_Berry}
\end{figure}

We show the energy band in the first-shell approximation in Fig.\ \ref{fig:SI_bands}(a) and including the 4th and 5th shell with opposite coefficients such that $\mathcal M_x$ is broken in Fig.\ \ref{fig:SI_bands}(b). The Berry curvature for the first shell and up to the second shell are shown in Fig.\ \ref{fig:SI_Berry}(a) and (b), respectively.

\subsection{First-shell approximation}

\begin{figure}
    \centering
    \includegraphics[width=0.5\linewidth]{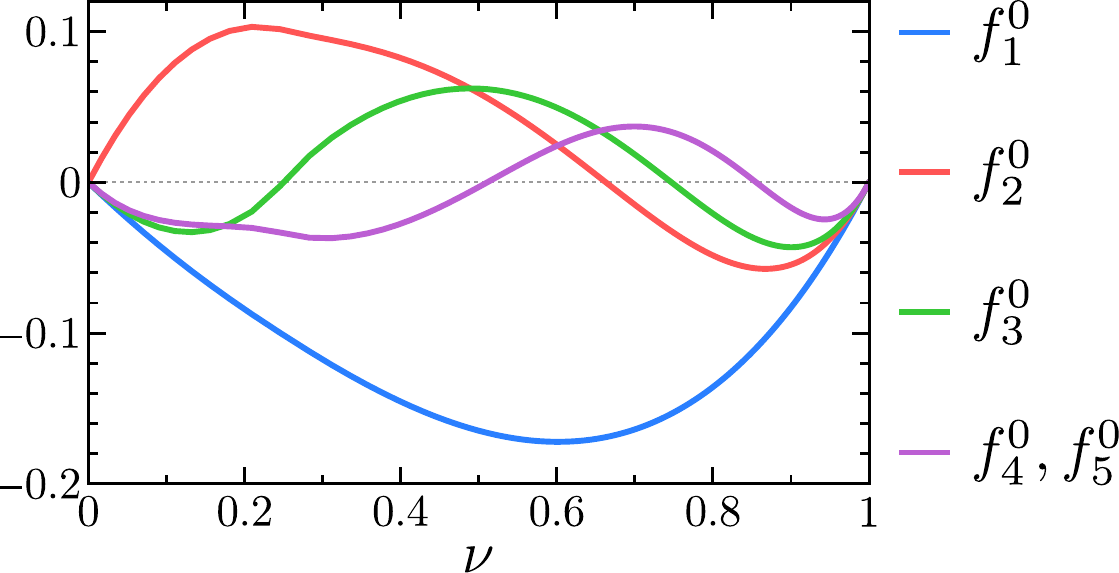}
    \caption{Fourier components of the Fermi function for $\varepsilon_j = \varepsilon_1 \delta_{j,1}$ and $k_BT/\varepsilon_1=0.01$ as a function of the filling $\nu$.}
    \label{fig:SI_Fermi}
\end{figure}

\begin{figure}
    \centering
    \includegraphics[width=0.9\linewidth]{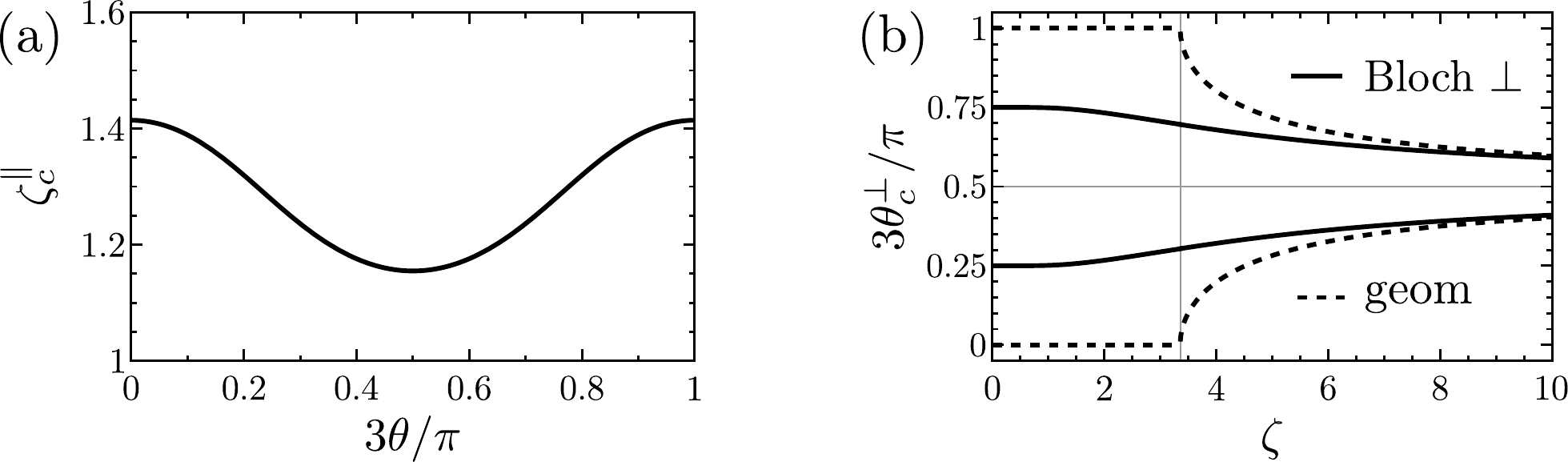}
    \caption{(a) Critical field where $F_\text{Bloch}^\parallel(\zeta,\theta)$ attains a maximum as a function of the field direction. (b) Critical angles where the absolute value of the transverse roses $F_\text{Bloch}^\perp(\zeta,\theta)$ and $F_\text{geom}(\zeta,\theta)$ reaches a maximum as a function of the field strength. Notice that the maxima both converge to $\theta=\pi/6$ for $\zeta \gg 1$, for which the transverse roses vanish by $\mathcal M_x$ or $\mathcal C_{2y}$ symmetry.}
    \label{fig:SI_JMax}
\end{figure}

We start by calculating the currents for the first shell, i.e., we set $\varepsilon_j = \varepsilon_1 \delta_{j,1}$ and $\Omega_j = \Omega_1 \delta_{j,1}$. In this case, the Bloch current can be written as
\begin{equation}
    \bm J_\text{Bloch} = \frac{e}{V_c \hbar} \frac{\varepsilon_1 f^0_1}{2} \sum_{n=1}^3 \left( \frac{i \bm L_n^{(1)}}{1 - ie \tau \bm E \cdot \bm L_n^{(1)} / \hbar} - \frac{i \bm L_n^{(1)}}{1 + ie \tau \bm E \cdot \bm L_n^{(1)} / \hbar} \right) = -\frac{e\varepsilon_1 f^0_1}{V_c \hbar} \sum_{n=1}^3 \frac{\bm L_n^{(1)} ( e \tau \bm E \cdot \bm L_n^{(1)} / \hbar )}{1 + ( e \tau \bm E \cdot \bm L_n^{(1)} / \hbar )^2},
\end{equation}
where $f_1^0$ is the Fourier component of the Fermi function for $\bm R = \pm \bm L_n^{(1)}$ ($n=1,2,3$) which are all real and equal due to $\mathcal T$ and $\mathcal C_{3z}$, respectively. Next, we introduce the dimensionless quantity
\begin{equation}
    \omega_B \tau = e \tau E L / \hbar,
\end{equation}
with $\omega_B$ the Bloch frequency and where $E = |\bm E|$. If we parameterize the electric field by an angle $\theta$ such that $\bm E = E \left( \cos \theta, \sin \theta \right)$, the component parallel to the field becomes
\begin{align}
    J_\text{Bloch}^\parallel \equiv \hat E \cdot \bm J_\text{Bloch} & = -\frac{eL\varepsilon_1 f^0_1}{V_c \hbar} \sum_{n=1}^3 \frac{\omega_B \tau ( \hat E \cdot \bm L_n^{(1)} / L )^2}{1 + ( \omega_B \tau )^2 ( \hat E \cdot \bm L_n^{(1)} / L )^2} \\
    & = -\frac{eL\varepsilon_1 f^0_1}{V_c \hbar} \sum_{n=1}^3 \frac{\omega_B \tau \cos^2 \theta_n}{1 + ( \omega _B\tau )^2 \cos^2\theta_n},
\end{align}
where we defined the angles $\theta_n$ through
\begin{equation}
    \hat E \cdot \bm L_n^{(1)} = L \cos \theta_n,
\end{equation}
with $\theta_1 = \theta - \pi/3$, $\theta_2 = \theta + \pi$, and $\theta_3 = \theta+\pi/3$. This yields
\begin{equation}
    J_\text{Bloch}^\parallel(\bm E) = -\frac{3eL\varepsilon_1 f^0_1}{V_c \hbar} \, F_\text{Bloch}^\parallel(\omega_B \tau,\theta), \qquad F_\text{Bloch}^\parallel(\zeta,\theta) = \zeta \, \frac{8 + 6 \zeta^2 + \zeta^4 \cos^2(3\theta)}{16 + 24 \zeta^2 + 9 \zeta^4 + \zeta^6 \cos^2(3\theta)}.
\end{equation}
Note that the filling of the band, as well as the effect of temperature, only enters via the overall factor $f_1^0$, which is plotted in Fig.\ \ref{fig:SI_Fermi}. The function $F_\text{Bloch}^\parallel$ for fixed field strength as a function of $\theta$ is called a \emph{rose curve}. It is shown in Fig.\ 2(b) of the main text. Because $F_\text{Bloch}^\parallel(\zeta,\theta)$ has no zeroes for finite $\zeta$, the rose only has a single petal. We further find that $\text{max} \, F_\text{Bloch}^\parallel \approx 0.3$ where the critical field is determined by a fifth-order polynomial in $\zeta^2$. For $\theta = m\pi/3$ ($m \in \mathds Z$) its real roots are $\zeta^2 = 2$ while for $\theta = \pi/6 + m\pi/3$ we find $\zeta^2 = 4/3$. This is shown in Fig.\ \ref{fig:SI_JMax}(a).
If we define the direction perpendicular to the electric field as $\hat E \times \hat z = \left( \sin \theta, -\cos \theta \right)$, we find $( \hat E \times \hat z ) \cdot \bm L_n^{(1)} = L \sin \theta_n$ such that
\begin{equation}
    J_\text{Bloch}^\perp(\bm E) \equiv ( \hat E \times \hat z ) \cdot \bm J_\text{Bloch} = - \frac{eL\varepsilon_1 f^0_1}{V_c \hbar} \sum_{n=1}^3 \frac{\omega_B \tau \sin \theta_n \cos \theta_n}{1 + ( \omega_B \tau )^2 \cos^2\theta_n} = - \frac{3eL\varepsilon_1 f^0_1}{V_c \hbar} \, F_\text{Bloch}^\perp(\omega_B \tau,\theta),
\end{equation}
with
\begin{equation}
    F_\text{Bloch}^\perp(\zeta,\theta) = \frac{\zeta^5 \sin(3\theta) \cos(3\theta)}{16 + 24 \zeta^2 + 9 \zeta^4 + \zeta^6 \cos^2(3\theta)}.
\end{equation}
A polar plot of $|F_\text{Bloch}^\perp(\zeta,\theta)|$ is shown in Fig.\ 2(c) of the main text for different values of $\zeta$. The transverse Bloch rose has twelve petals, since $F_\text{Bloch}^\perp$ vanishes for $\theta = m \pi / 6$ ($m \in \mathds Z$). Three of these angles correspond to the three mirror axes where the transverse response vanishes. The other three angles are a consequence of $\mathcal T$ in combination with mirror symmetry. Consider, for example, the case $E_y = 0$ (other axes are obtained by $\mathcal C_{3z}$). In this case, time-reversal symmetry dictates that the Bloch current is odd in $E_x$ while mirror symmetry requires that the transverse component is even in $E_x$, and therefore $J_\text{Bloch}^\perp$ vanishes for $E_y = 0$. Summarized, we see that in the presence of time-reversal symmetry, the transverse Bloch current vanishes when the electric field is either parallel or perpendicular to a mirror axis. The extremal angles (tips of the petals) of $F_\text{Bloch}^\perp(\zeta,\theta)$ are shown as a function of $\zeta$ in Fig.\ \ref{fig:SI_JMax}(b).

For the geometric current, we have $\hat E \cdot \bm J_\text{geom} = 0$ and in the first-shell approximation,
\begin{align}
    J_\text{geom}(\bm E) \equiv ( \hat E \times \hat z ) \cdot \bm J_\text{geom} & = - \frac{e^2E}{V_c \hbar} \frac{\Omega_1 f_1^0}{2} \sum_{n=1}^3 \left( \frac{i}{1 - i e \tau \bm E \cdot \bm L_n^{(1)} / \hbar} - \frac{i}{1 + i e \tau \bm E \cdot \bm L_n^{(1)} / \hbar} \right) \\
    & = \frac{e^2E}{V_c \hbar} \frac{\Omega_1 f_1^0}{2} \sum_{n=1}^3 \frac{2e \tau \bm E \cdot \bm L_n^{(1)} / \hbar}{1 + ( e \tau \bm E \cdot \bm L_n^{(1)} / \hbar )^2} \\
    & = \frac{eL}{V_c \tau} \frac{\Omega_1f_1^0}{L^2} \, \sum_{n=1}^3 \frac{( \omega_B \tau )^2 \cos \theta_n}{1 + ( \omega_B \tau )^2 \cos^2 \theta_n} = \frac{3eL}{V_c \tau} \frac{\Omega_1f_1^0}{L^2} F_\text{geom}(\omega_B \tau,\theta),
\end{align}
with
\begin{equation}
    F_\text{geom}(\zeta,\theta) = \frac{\zeta^4 \left( 4 + \zeta^2 \right) \cos(3\theta)}{16 + 24 \zeta^2 + 9 \zeta^4 + \zeta^6 \cos^2(3\theta)},
\end{equation}
whose absolute value is shown in Fig.\ 2(d) of the main text. The geometric rose has six petals, since mirror symmetry precludes a transverse response when the electric field lies along a mirror axis. We find that minima of $|F_\text{geom}(\zeta,\theta)|$ always occur at $\theta = m \pi/3$ and extrema occur for
\begin{equation}
    \sin(6\theta) = 0, \qquad \cos^2(3\theta) = \left( \frac{4+3\zeta^2}{\zeta^3} \right)^2,
\end{equation}
such that the right-hand side has to be smaller or equal to one which yields $\zeta \geq 3.355$ approximately. For $\zeta \gg 1$, the maxima converge to $\theta = \pi/6 + m \pi/3$. However, exactly at these angles, which correspond to mirror axes, the geometric current vanishes. The extremal angles of $F_\text{geom}(\zeta,\theta)$ are shown in Fig.\ \ref{fig:SI_JMax}(b) as a function of $\zeta$.

\paragraph{Weak-field limit}

The weak-field ($\zeta \ll 1$) expansions are given by
\begin{align}
    F_\text{Bloch}^\parallel(\zeta,\theta) & \simeq \frac{\zeta}{2} - \frac{3\zeta^3}{8} + \frac{10 + \cos(6\theta)}{32} \, \zeta^5, \\
    F_\text{Bloch}^\perp(\zeta,\theta) & \simeq \frac{\sin(3\theta) \cos(3\theta)}{16} \, \zeta^5, \\
    F_\text{geom}(\zeta,\theta) & \simeq \frac{\cos(3\theta)}{4} \, \zeta^4,
\end{align}
consistent with the symmetry analysis. The transverse Bloch current only appears at fifth order because the linear and cubic terms are forbidden by $\mathcal C_{3z}$. Indeed, for any in-plane vector $\bm R$, the longitudinal and transverse components of the Bloch current contain the sums 
\begin{align}
    \sum_{n=1}^3 ( \hat E \cdot \mathcal C_{3z}^n \bm R )^2 & = \frac{3}{2} |\bm R|^2, \\
    \sum_{n=1}^3 [ ( \hat E \times \hat z ) \cdot \mathcal C_{3z}^n \bm R ] ( \hat E \cdot \mathcal C_{3z}^n \bm R ) & = 0,    
\end{align}
such that the linear term in the transverse Bloch current is forbidden. Otherwise it would result in a symmetric part of the transverse linear conductivity. Similarly, for the cubic terms
\begin{align}
    \sum_{n=1}^3 ( \hat E \cdot \mathcal C_{3z}^n \bm R )^4 & = \frac{9}{8} |\bm R|^4, \\
    \sum_{n=1}^3 [ ( \hat E \times \hat z ) \cdot \mathcal C_{3z}^n \bm R ] ( \hat E \cdot \mathcal C_{3z}^n \bm R )^3 & = 0,   
\end{align}
while all high-order terms are generally nonzero and depend on $\hat E$. We also note that the lowest-order geometric current is quartic in the field. The quadratic term, correspodning to the Berry curvature dipole, which is allowed by time-reversal symmetry, is proportional to
\begin{equation}
    \hat E \cdot \sum_{\bm R} \bm R f_{\bm R}^0 \Omega_{-\bm R},
\end{equation}
where
\begin{equation}
    \sum_{\bm R} \bm R f_{\bm R}^0 \Omega_{-\bm R} = \sum_{\bm R}  \bm R f_{\mathcal C_{3z} \bm R}^0 \Omega_{-\mathcal C_{3z} \bm R} = \mathcal C_{3z}^{-1} \sum_{\bm R} \bm R f_{\bm R}^0 \Omega_{-\bm R},
\end{equation}
such that the vector sum vanishes. The cubic term involving the Berry curvature quadrupole, is forbidden by time-reversal symmetry. This is true for all odd powers:
\begin{equation}
    \left( \hat z \times \bm E \right) \sum_{\bm R} \left( \bm E \cdot \bm R \right)^{2n} f_{\bm R}^0 \Omega_{-\bm R} \propto \sum_{\bm R} \left( -\bm E \cdot \bm R \right)^{2n} f_{-\bm R}^0 \Omega_{\bm R} \overset{\mathcal T}{=} -\sum_{\bm R} \left( \bm E \cdot \bm R \right)^{2n} f_{\bm R}^0 \Omega_{-\bm R}.
\end{equation}

\paragraph{Strong-field limit}

The strong-field ($\zeta \gg 1$) expansions are given by
\begin{align}
    F_\text{Bloch}^\parallel(\zeta,\theta) & \simeq \frac{1}{\zeta}, \\
    F_\text{Bloch}^\perp(\zeta,\theta) & \simeq \frac{\tan(3\theta)}{\zeta}, \\
    F_\text{geom}(\zeta,\theta) & \simeq \frac{1}{\cos(3\theta)},
\end{align}
where the last two lines hold only for $\theta \neq \pi/6 + m \pi/3$. Precisely at these angles, the transverse currents vanish because of mirror symmetry, which is conserved if we only include the first shell.
\begin{table}
\renewcommand{\arraystretch}{1.5}
\setlength{\tabcolsep}{5pt}
\caption{Angles and scaling factors of the coordination shells of the triangular lattice, which are regular hexagons with radius $L_j$ rotated by an angle $\theta_j$ relative to the first shell, shown here up to the seventh shell.}
\begin{center}
\begin{tabular}{l c c c c c c c}
\Xhline{1pt}
Shell & 1 & 2 & 3 & 4 & 5 & 6 & 7 \\
\Xhline{1pt}
$L_j/L$ & $1$ & $\sqrt{3}$ & $2$ & $\sqrt{7}$ & $\sqrt{7}$ & $3$ & $2\sqrt{3}$ \\
\hline
$\theta_j$ & $0$ & $\pi/6$ & $0$ & $\pi/6 + \arctan \left( \sqrt{3}/5 \right)$ & $\pi/6 - \arctan \left( \sqrt{3}/5 \right)$ & $0$ & $\pi/6$ \\
\Xhline{1pt}
\end{tabular}
\label{tab:SIshells}
\end{center}
\end{table}

\subsection{General case including all shells}

To obtain the general expression including contributions from all shells, we first note that all higher-order shells are obtained from the first shell by a rotation and a scaling. Hence, the results obtained for the first shell can be used to find the contribution of any shell. For example, the result for the second shell is obtained by sending
\begin{align}
    L & \mapsto \sqrt{3} \, L, \\
    \zeta & \mapsto \sqrt{3} \, \zeta, \\
    \theta & \mapsto \theta + \frac{\pi}{6},
\end{align}
in the first-shell expressions for $\bm J_\text{Bloch}$ and $\bm J_\text{geom}$. In this way, we find
\begin{align}
    J_\text{Bloch}^\parallel(\bm E) & = -\frac{3eL}{V_c \hbar} \sum_j \frac{\varepsilon_j f_j^0 L_j}{L} \, F_\text{Bloch}^\parallel \left( \omega_B \tau L_j / L , \theta + \theta_j \right), \\
    J_\text{Bloch}^\perp(\bm E) & = -\frac{3eL}{V_c \hbar} \sum_j \frac{\varepsilon_j f_j^0 L_j}{L} \, F_\text{Bloch}^\perp \left( \omega_B \tau L_j / L , \theta + \theta_j \right), \\
    J_\text{geom}(\bm E) & = \frac{3eL}{V_c \tau} \sum_j \frac{\Omega_j f_j^0}{L^2} \frac{L}{L_j} \, F_\text{geom} \left( \omega_B \tau L_j / L , \theta + \theta_j \right),
\end{align}
with $\omega_B = eEL/\hbar$ and where the sums run over shells. Here $\varepsilon_j$ and $\Omega_j$ are the coefficients in the shell expansion of the band dispersion and the Berry curvature, respectively, and $f_j^0$ are the corresponding Fourier components of the Fermi function. An overview of the angles and the scaling factors up to the seventh shell is shown in Table \ref{tab:SIshells}.

\subsection{Differential conductance}

\begin{figure}
    \centering
    \includegraphics[width=\linewidth]{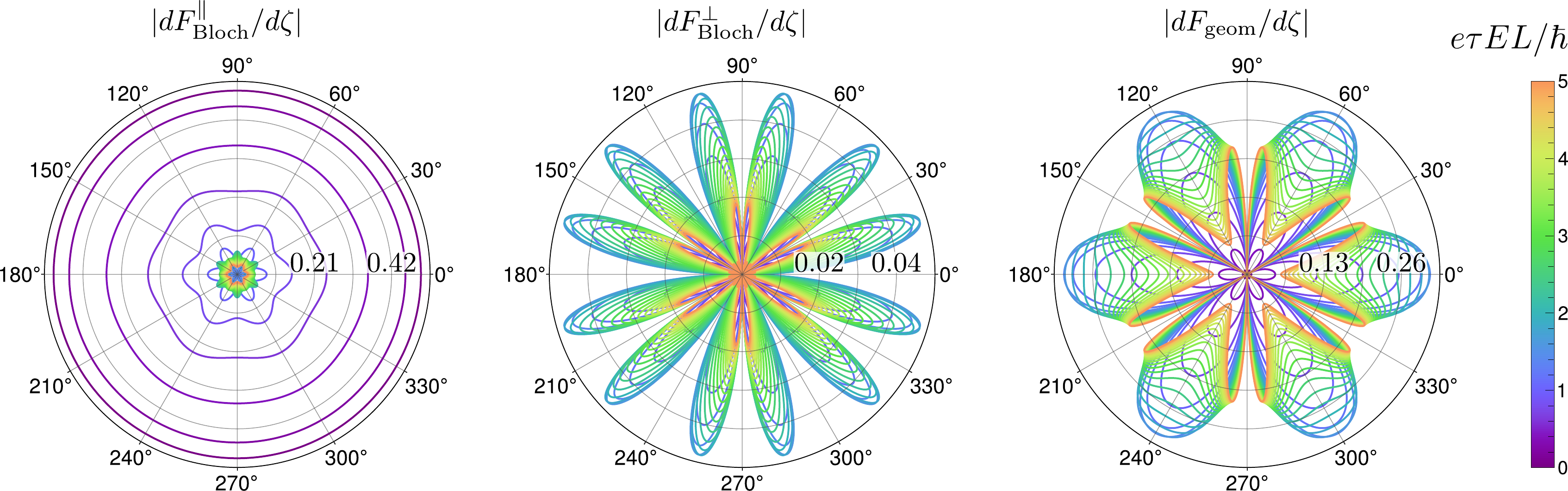}
    \caption{Differential conductance roses for the first shell.}
    \label{fig:SI_droses}
\end{figure}

We define the differential conductances as
\begin{align}
    \frac{d J_\text{Bloch}^\parallel}{dE} & = - \frac{e^2 \tau}{V_c \hbar^2} \sum_{\bm R} \frac{( \hat E \cdot \bm R )^2 f_{\bm R}^0 \varepsilon_{-\bm R}}{\left( 1 - i e \tau \bm E \cdot \bm R / \hbar \right)^2}, \\
    \frac{d J_\text{Bloch}^\perp}{dE} & = - \frac{e^2 \tau}{V_c \hbar^2} \sum_{\bm R} \frac{[ ( \hat E \times \hat z ) \cdot \bm R ] ( \hat E \cdot \bm R ) f_{\bm R}^0 \varepsilon_{-\bm R}}{\left( 1 - i e \tau \bm E \cdot \bm R / \hbar \right)^2}, \\
    \frac{d J_\text{geom}}{dE} & = - \frac{e^2}{V_c \hbar} \sum_{\bm R} \frac{f_{\bm R}^0 \Omega_{-\bm R}}{\left( 1 - i e \tau \bm E \cdot \bm R / \hbar \right)^2},
\end{align}
and thus
\begin{align}
    \frac{dJ_\text{Bloch}^\parallel}{dE}(\bm E) & = -\frac{3e^2\tau L^2}{V_c \hbar^2} \sum_j \frac{\varepsilon_j f_j^0 L_j}{L} \left. \frac{dF_\text{Bloch}^\parallel \left( \zeta L_j / L , \theta + \theta_j \right)}{d\zeta} \right|_{\zeta=\omega_B\tau}, \\
    \frac{dJ_\text{Bloch}^\perp}{dE}(\bm E) & = -\frac{3e^2\tau L^2}{V_c \hbar^2} \sum_j \frac{\varepsilon_j f_j^0 L_j}{L} \left. \frac{dF_\text{Bloch}^\perp \left( \zeta L_j / L , \theta + \theta_j \right)}{d\zeta} \right|_{\zeta=\omega_B\tau}, \\
    \frac{dJ_\text{geom}}{dE}(\bm E) & = \frac{3e^2L^2}{V_c \hbar} \sum_j \frac{\Omega_j f_j^0}{L^2} \frac{L}{L_j} \left. \frac{dF_\text{geom} \left( \zeta L_j / L , \theta + \theta_j \right)}{d\zeta} \right|_{\zeta=\omega_B\tau}.
\end{align}
We show the differential conductance roses for the first shell in Fig.\ \ref{fig:SI_droses}.

\subsection{Summing contributions from two valleys}

Finally, we consider two decoupled energy bands that are isolated from other bands with band dispersion $\varepsilon_{\bm k}^\nu$ and Berry curvature $\Omega_{\bm k}^\nu$, and that are related by time-reversal symmetry. Here, $\nu = \pm 1$ is the valley index. Time-reversal symmetry implies a relation between the energy bands and the Berry curvature of the two valleys:
\begin{equation}
    \varepsilon_{\bm k}^\nu = \varepsilon_{-\bm k}^{-\nu}, \qquad \Omega_{\bm k}^\nu = -\Omega_{-\bm k}^{-\nu}.
\end{equation}
Similarly in real space,
\begin{align}
    \varepsilon_{\bm R}^\nu & = \varepsilon_{-\bm R}^{-\nu} = \left( \varepsilon_{\bm R}^{-\nu} \right)^*, \\
    \Omega_{\bm R}^\nu & = -\Omega_{-\bm R}^{-\nu} = - \left( \Omega_{\bm R}^{-\nu} \right)^*.
\end{align}
Let us consider the specific case where the symmetries of a single valley are given by the magnetic point group $3m' =\left< \mathcal C_{3z}, \mathcal M_x \mathcal T \right>$. As before, we expand the energy bands in terms of the coordination shells:
\begin{align}
    \varepsilon_{\bm k}^\nu & = \varepsilon_0 + \varepsilon_1 \sum_{n=1}^3 \cos ( \bm k \cdot \bm L_n^{(1)} + \nu  \phi ) + \cdots, \\
    \Omega_{\bm k}^\nu & = \nu \Omega_0 + \Omega_1 \sum_{n=1}^3 \sin ( \bm k \cdot \bm L_n^{(1)} + \nu \xi ) + \cdots.   
\end{align}
where $\phi$ and $\xi$ are phases that are allowed because $\mathcal T$ is broken within a single valley. Using the relations between the real-space Fourier components of different valleys, the total current can be written solely in terms of quantities at a single valley,
\begin{align}
    \bm J_\text{Bloch} & = \frac{2e}{V_c \hbar} \sum_{\bm R} \frac{i \bm R \, \text{Re}\left( f_{\bm R}^{0+} \varepsilon_{-\bm R}^+ \right)}{1 - i e \tau \bm E \cdot \bm R / \hbar}, \\
    \bm J_\text{geom} & = \left( \hat z \times \bm E \right) \frac{2e^2}{V_c \hbar} \sum_{\bm R} \frac{i \text{Im} \left( f_{\bm R}^{0+} \Omega_{-\bm R}^+ \right)}{1 - i e \tau \bm E \cdot \bm R / \hbar}.
\end{align}
In the first-shell approximation, we find
\begin{align}
    \bm J_\text{Bloch}(\bm E) & = - \frac{6e L \varepsilon_1 |f_1^0| \cos(\chi - \phi)}{V_c \hbar} \left[ \hat E F_\text{Bloch}^\parallel(\omega_B \tau,\theta) + ( \hat E \times \hat z ) \, F_\text{Bloch}^\perp(\omega_B \tau,\theta) \right], \\
    \bm J_\text{geom}(\bm E) & = ( \hat E \times \hat z ) \, \frac{6eL}{V_c\tau} \frac{\Omega_1 |f_1^0| \cos(\chi-\xi)}{L^2} \, F_\text{geom}(\omega_B \tau,\theta),
\end{align}
where $f_{\pm \bm R}^{0\nu} = |f_1^0| e^{\pm i\nu\chi}$ for $\bm R = \bm L_n^{(1)}$ ($n=1,2,3$).

\section{Periodically-buckled graphene}

\subsection{Strain profile from height modulation}

Given a height modulation of the monolayer graphene, induced by a buckling transition, we want to obtain the corresponding strain tensor. The strain tensor $u_{ij}$ ($i,j=x,y$) is defined \cite{Landau1970} by considering the change in length between two points with initial (infinitesimal and in-plane) separation $dr_i$ after a deformation: $\left[ (dr_i + du_i)^2 + dh^2 \right ] - dr_i^2 \equiv 2u_{ij} dr_i dr_j$. Up to lowest order in the displacements, the strain tensor is given by
\begin{equation}
    u_{ij}(\bm r) = \frac{1}{2} \left[ \partial_i u_j + \partial_j u_i + \left( \partial_i h \right) \left( \partial_j h \right) \right],
\end{equation}
with $\partial_i = \partial/\partial r_i$ and where $u_i(\bm r)$ and $h(\bm r)$ are the in-plane and out-of-plane displacements, respectively.  

If the displacements are periodic, we can write them as a Fourier series:
\begin{align}
    u_i(\bm r) & = \sum_{\G} u_{i\G} \, e^{i \G \cdot \bm r}, \\
    h(\bm r) & = \sum_{\G} h_{\G} \, e^{i \G \cdot \bm r},
\end{align}
where $\G$ is a reciprocal lattice vector of the periodic modulation (not of the monolayer graphene) and $u_{i\bm 0} = h_{\bm 0} = 0$. For later convenience, we also define
\begin{equation}
        f_{ij}(\bm r) \equiv \left[ \partial_i h(\bm r) \right] \left[ \partial_j h(\bm r) \right] = \sum_{\G} f_{ij\G} \, e^{i \G \cdot \bm r},
\end{equation}
where
\begin{equation}
    f_{ij\G} = - \sum_{\G'} h_{\G'} h_{\G - {\G}'} \mathcal G_i' \left( \mathcal G_j - \mathcal G_j'\right).
\end{equation}
The strain tensor becomes
\begin{equation}
    u_{ij}(\bm r) = \frac{1}{2} \sum_{\G} \left[ i \left( \mathcal G_i u_{j\G} + \mathcal G_j u_{i\G} \right) + f_{ij\G} \right] e^{i \G \cdot \bm r}.
\end{equation}
A fixed height profile $h(\bm r)$ will give rise to in-plane displacements as the graphene lattice relaxes. The in-plane displacements can be found by minimizing the elastic energy density \cite{Guinea2008,Phong2022}:
\begin{align}
    \mathcal E_\text{elas} & = \frac{1}{V} \int d^2 \bm r \left[ \frac{\lambda}{2} \left( u_{xx} + u_{yy} \right)^2 + \mu \left( u_{xx}^2 + u_{yy}^2 + 2u_{xy}^2 \right) \right] \\
    & = \frac{1}{V} \int d^2 \bm r \left[ \left( \frac{\lambda}{2} + \mu \right) \left( u_{xx}^2 + u_{yy}^2 \right) + \lambda u_{xx} u_{yy} + 2 \mu u_{xy}^2 \right],
\end{align}
where $\lambda$ and $\mu$ are the Lam\'e parameters for graphene. Plugging in the Fourier expansions, we obtain
\begin{align}
     \frac{1}{V} \int d^2 \bm r \, u_{ii}^2 & = \frac{1}{V} \sum_{\G,\G'} \int d^2 \bm r \left( i \mathcal G_i u_{i\G} + \frac{f_{ii \G}}{2} \right) \left( i \mathcal G_i' u_{i\G'} + \frac{f_{ii \G'}}{2} \right) e^{i ( \G + \G' ) \cdot \bm r} \\
     & = \sum_{\G} \left| i \mathcal G_i u_{i\G} + \frac{f_{ii \G}}{2} \right|^2, \\
     \frac{1}{V} \int d^2 \bm r \, u_{xx} u_{yy} & = \sum_{\G} \left( i\mathcal G_x u_{x\G} + \frac{f_{xx \G}}{2} \right) \left( -i\mathcal G_y u_{y\G}^* + \frac{f_{yy \G}^*}{2} \right) \\
     & = \frac{1}{2} \sum_{\G} \left[ \left( i\mathcal G_x u_{x\G} + \frac{f_{xx \G}}{2} \right) \left( -i\mathcal G_y u_{y\G}^* + \frac{f_{yy \G}^*}{2} \right) + \text{c.c.} \right], \\
     \frac{1}{V} \int d^2 \bm r \, u_{xy}^2 & = \frac{1}{4} \sum_{\G} \left( i\mathcal G_x u_{y\G} + i\mathcal G_y u_{x\G} + f_{xy\G} \right) \left( -i\mathcal G_x u_{y\G}^* - i\mathcal G_y u_{x\G}^* + f_{xy\G}^* \right).
\end{align}
Hence, the elastic energy density becomes
\begin{align}
    \mathcal E_\text{elas} & = \left( \frac{\lambda}{2} + \mu \right) \sum_{\G} \left( i\mathcal G_x u_{x\G} + \frac{f_{xx \G}}{2} \right) \left( -i\mathcal G_x u_{x\G}^* + \frac{f_{xx \G}^*}{2} \right) \\
    & + \left( \frac{\lambda}{2} + \mu \right) \sum_{\G} \left( i\mathcal G_y u_{y\G} + \frac{f_{yy \G}}{2} \right) \left( -i\mathcal G_y u_{y\G}^* + \frac{f_{yy \G}^*}{2} \right) \\
    & + \frac{\lambda}{2} \sum_{\G} \left[ \left( i\mathcal G_x u_{x\G} + \frac{f_{xx\G}}{2} \right) \left( -i\mathcal G_y u_{y\G}^* + \frac{f_{yy\G}^*}{2} \right) + \text{c.c.} \right] \\
    & + \frac{\mu}{2} \sum_{\G} \left( i\mathcal G_x u_{y\G} + i\mathcal G_y u_{x\G} + f_{xy\G} \right) \left( -i\mathcal G_x u_{y\G}^* - i \mathcal G_y u_{x\G}^* + f_{xy\G}^* \right).
\end{align}
By extremizing the elastic energy with respect to $u_{i\G}^*$ we obtain equations for the Fourier components $u_{i\G}$ in terms of $f_{ij\G}$ (and thus $h_{i\G}$). We find
\begin{align}
    \frac{\partial \mathcal E}{\partial u_{x\G}^*} & = -i \mathcal G_x \left[ \left( \frac{\lambda}{2} + \mu \right) \left( i\mathcal G_x u_{x\G} + \frac{f_{xx \G}}{2} \right) + \frac{\lambda}{2} \left( i\mathcal G_y u_{y\G} + \frac{f_{yy \G}}{2} \right) \right] - i \mathcal G_y \frac{\mu}{2} \left( i\mathcal G_x u_{y\G} + i \mathcal G_y u_{x\G} + f_{xy\G} \right), \\
    \frac{\partial \mathcal E}{\partial u_{y\G}^*} & = -i \mathcal G_y \left[ \left( \frac{\lambda}{2} + \mu \right) \left( i\mathcal G_y u_{y\G} + \frac{f_{yy \G}}{2} \right) + \frac{\lambda}{2} \left( i\mathcal G_x u_{x\G} + \frac{f_{xx \G}}{2} \right) \right] - i \mathcal G_x \frac{\mu}{2} \left( i\mathcal G_x u_{y\G} + i \mathcal G_y u_{x\G} + f_{xy\G} \right).
\end{align}
Setting the above two equations equal to zero, yields solutions
\begin{align}
    u_{x\G} & = \frac{i}{2 \left( \lambda + 2 \mu \right) |\G|^4} \left\{ f_{xx}^{\G} \mathcal G_
    x \left[ \mathcal G_x^2 \left( \lambda + 2 \mu \right) + \mathcal G_y^2 \left( 3 \lambda + 4 \mu \right) \right] + \left( f_{yy}^{\G} \mathcal G_x - 2 f_{xy}^{\G} \mathcal G_y \right) \left[ \mathcal G_x^2 \lambda - \mathcal G_y^2 \left( \lambda + 2 \mu \right) \right]  \right\}, \\
    u_{y\G} & = \frac{i}{2 \left( \lambda + 2 \mu \right) |\G|^4} \left\{ f_{yy}^{\G} \mathcal G_
    y \left[ \mathcal G_y^2 \left( \lambda + 2 \mu \right) + \mathcal G_x^2 \left( 3 \lambda + 4 \mu \right) \right] + \left( f_{xx}^{\G} \mathcal G_y - 2 f_{xy}^{\G} \mathcal G_x \right) \left[ \mathcal G_y^2 \lambda - \mathcal G_x^2 \left( \lambda + 2 \mu \right) \right]  \right\}.
\end{align}

\subsubsection{Pseudomagnetic field}

Shear strain breaks the microscopic $\mathcal C_{3z}$ symmetry and couples to the low-energy Dirac electrons of graphene through a pseudo vector potential $\nu \bm{\mathcal A}(\bm r)$ with $\nu = \pm 1$ the valley index and \cite{Suzuura2005,Manes2007,Guinea2010,Vozmediano2010,deJuan2013}
\begin{equation} \label{eq:pA}
    \bm{\mathcal A} = -\frac{\sqrt{3} \hbar \beta}{2ea} \begin{pmatrix} u_{xx} - u_{yy} \\ - 2 u_{xy} \end{pmatrix},
\end{equation}
where $e > 0$ is the elementary charge, $a \approx 0.246$~nm is the lattice constant of graphene, and $\beta \sim 1$ is the electron Gr\"uneisen parameter for graphene. By using the results given above, we find that
\begin{align}
    u_{xx}^{\G} - u_{yy}^{\G} & = \frac{\left( \lambda + \mu \right) \left( \mathcal G_y^2 - \mathcal G_x^2 \right) \left( \mathcal G_x^2 f_{yy}^{\G} - 2 \mathcal G_x \mathcal G_y f_{xy}^{\G} + \mathcal G_y^2 f_{xx}^{\G} \right)}{\left( \lambda + 2 \mu \right) |\G|^4}, \\
    -2u_{xy}^{\G} & = \frac{2 \left( \lambda + \mu \right) \mathcal G_x \mathcal G_y \left( \mathcal G_x^2 f_{yy}^{\G} - 2 \mathcal G_x \mathcal G_y f_{xy}^{\G} + \mathcal G_y^2 f_{xx}^{\G} \right)}{\left( \lambda + 2 \mu \right) |\G|^4},
\end{align}
such that
\begin{equation}
    \bm{\mathcal A}(\bm r) = \frac{\sqrt{3} \hbar \beta}{2ea} \frac{\lambda + \mu}{\lambda + 2 \mu} \sum_{\G} \frac{\mathcal G_x^2 f_{yy}^{\G} - 2 \mathcal G_x \mathcal G_y f_{xy}^{\G} + \mathcal G_y^2 f_{xx}^{\G}}{|\G|^4} \begin{pmatrix} \mathcal G_x^2 - \mathcal G_y^2 \\ -2 \mathcal G_x \mathcal G_y
    \end{pmatrix} e^{i \G \cdot \bm r}.
\end{equation}
Likewise, the pseudomagnetic field $\bm{\mathcal B}(\bm r) = \mathcal B(\bm r) \hat z$ becomes
\begin{equation}
    \mathcal B(\bm r) = \partial_x \mathcal A_y - \partial_y \mathcal A_x = \frac{\sqrt{3} \hbar \beta}{2ea} \frac{\lambda + \mu}{\lambda + 2 \mu} \sum_{\G} \frac{i \mathcal G_y \left( \mathcal G_y^2 - 3 \mathcal G_x^2 \right) \left( \mathcal G_x^2 f_{yy}^{\G} - 2 \mathcal G_x \mathcal G_y f_{xy}^{\G} + \mathcal G_y^2 f_{xx}^{\G} \right)}{|\G|^4} \, e^{i \G \cdot \bm r}.
\end{equation}
Notice that $\G = \bm 0$ does not contribute (i.e., there is no net flux) since the nominator scales as $|\G|^5$. 

\subsubsection{Triangular height profile}

We now consider a height profile that conserves $\mathcal C_{3z}$ symmetry in the first-star approximation,
\begin{equation}
    h(\bm r) = h_0 \sum_{n=1}^3 \cos \left( \G_n \cdot \bm r + \frac{\pi}{4} + \phi \right), \label{eq:SIht}
\end{equation}
where $\phi$ is a parameter that controls the shape of the height profile. Note that while $\phi$ cannot be absorbed in a coordinate shift, Eq.\ \eqref{eq:SIht} is invariant under $\phi \mapsto \phi + 2\pi/3$ up to an overall translation. The finite Fourier components are $h_{\pm \G_n} = h_0 e^{\pm i \left( \phi + \pi/4 \right)}/2$ ($n=1,2,3$) and where
\begin{equation}
    \G_1 = \frac{4\pi}{\sqrt{3}L} \begin{pmatrix} 0 \\ 1 \end{pmatrix}, \qquad \G_{2/3} = \frac{4\pi}{\sqrt{3}L} \begin{pmatrix} \mp \sqrt{3}/2 \\ -1/2 \end{pmatrix},
\end{equation}
with $\G_3 = - \G_1 - \G_2$ and $L$ the lattice constant of the height modulation. These are the three shortest nonzero reciprocal lattice vectors that are related by $\mathcal C_{3z}$ symmetry. We calculate all the Fourier components of $f_{ij}(\bm r)$ with Mathematica. We then find that
\begin{equation}
    \begin{aligned}
    & \frac{i \mathcal G_y \left( \mathcal G_y^2 - 3 \mathcal G_x^2 \right) \left( \mathcal G_x^2 f_{yy}^{\G} - 2 \mathcal G_x \mathcal G_y f_{xy}^{\G} + \mathcal G_y^2 f_{xx}^{\G} \right)}{|\G|^4} \\
    & = \frac{3\mathcal G^3h_0^2}{8} \left[ e^{-2i\phi} \left( \delta_{\G,\G_1} + \delta_{\G,\G_2} + \delta_{\G,\G_3} \right) + e^{2i\phi} \left( \delta_{\G,-\G_1} + \delta_{\G,-\G_2} + \delta_{\G,-\G_3} \right) \right],
    \end{aligned}
\end{equation}
with $\mathcal G=4\pi/\sqrt{3}L$ and therefore
\begin{equation}
    \mathcal B(\bm r) = \mathcal B_0 \sum_{n=1}^3 \cos \left( \G_n \cdot \bm r - 2\phi \right), \qquad \mathcal B_0 = \frac{\hbar \beta}{ea} \frac{\lambda + \mu}{\lambda + 2 \mu} \frac{8\pi^3h_0^2}{L^3}.
\end{equation}
We find that the pseudomagnetic field is invariant under $\phi \mapsto \phi + \pi/3$ up to an overall translation, which changes the sign of the height profile. Hence we can restrict ourselves to $\phi \in (-\pi/6,\pi/6]$. The pseudomagnetic field has $\mathcal C_{6z}$ symmetry for the special case $\phi = \pi/12$.

\subsubsection{Rectangular height profile}

Let us also consider a height profile that conserves $\mathcal C_{4z}$ symmetry in the first-star approximation,
\begin{equation}
    h(\bm r) = h_0 \sum_{n=1}^2 \cos \left( \G_n \cdot \bm r \right).
\end{equation}
In this case, a phase factor can always be absorbed in a coordinate shift since there are only two reciprocal lattice vectors. The finite Fourier components are now given by $h_{\pm \G_n} = h_0/2$ ($n=1,2$) where
\begin{equation}
    \G_1 = \frac{2\pi}{L} \begin{pmatrix} 1 \\ 0 \end{pmatrix}, \qquad \G_2 = \frac{2\pi}{L} \begin{pmatrix} 0 \\ 1 \end{pmatrix}.
\end{equation}
We calculate all the Fourier components of $f_{ij}(\bm r)$ with Mathematica and find
\begin{equation}
    \begin{aligned}
    & \frac{i \mathcal G_y \left( \mathcal G_y^2 - 3 \mathcal G_x^2 \right) \left( \mathcal G_x^2 f_{yy}^{\G} - 2 \mathcal G_x \mathcal G_y f_{xy}^{\G} + \mathcal G_y^2 f_{xx}^{\G} \right)}{|\G|^4} \\
    & = \frac{\mathcal G^3h_0^2}{4i} \left[ \left( \delta_{\G,\G_2+\G_1} + \delta_{\G,\G_2 - \G_1} \right) - \left( \delta_{\G,-\G_2 - \G_1} + \delta_{\G,-\G_2 + \G_1} \right) \right],
    \end{aligned}
\end{equation}
with $\mathcal G=2\pi/L$ and
\begin{align}
    \mathcal B(\bm r) = \mathcal B_0 \left( \sin \left[ \left( \G_2 + \G_1 \right) \cdot \bm r \right] + \sin \left[ \left( \G_2 - \G_1 \right) \cdot \bm r \right] \right), \qquad \mathcal B_0 = \frac{\hbar \beta}{ea} \frac{\lambda + \mu}{\lambda + 2 \mu} \frac{2 \sqrt{3} \pi^3h_0^2}{L^3}.
\end{align}
Under $\mathcal C_{2z}$, the pseudomagnetic field picks up an extra sign because the valleys are interchanged. Hence the pseudomagnetic field for a height profile with $\mathcal C_{4z}$ symmetry always has $\mathcal C_{2z}$ symmetry, as expected. In this case, band crossings between backfolded bands are protected locally in momentum space by $\mathcal C_{2z}\mathcal T$ symmetry.

\subsection{Continuum model}

When the height profile varies slowly compared to the graphene lattice constant, i.e., $L \gg a$, we can use the valley-projected continuum theory \cite{Phong2022,DeBeule2023},
\begin{equation}
    \hat H_\nu = \int d^2 \bm r \, \hat \psi_\nu^\dag(\bm r) \left\{ \hbar v \left[ -i\nabla + \frac{\nu e}{\hbar} \,  \bm{\mathcal A}(\bm r) \right] \cdot \left( \nu \sigma_x, \sigma_y \right) + \mathcal V(\bm r) \sigma_0 \right\} \hat \psi_\nu(\bm r),
\end{equation}
where $\sigma_x$ and $\sigma_y$ are Pauli matrices, $\sigma_0$ is the $2\times2$ identity matrix, $\hat \psi_\nu = (\hat \psi_{\nu A}, \hat \psi_{\nu B})^t$ are field operators satisfying $\{\hat \psi_\nu^\dag(\bm r),\hat \psi_{\nu'}(\bm r')\} = \delta_{\nu\nu'} \delta^{(2)}(\bm r -\bm r')$, and we take $\hbar v \approx 575.2 \; \text{meV} \, \text{nm}$ \cite{Castro2009}. The pseudo vector potential and scalar potential are given in terms of their Fourier series,
\begin{equation}
    \bm{\mathcal A}(\bm r) = \sum_{\G} \bm{\mathcal A}_{\G} \, e^{i \G \cdot \bm r}, \qquad \mathcal V(\bm r) = \mathcal V_0 \, \frac{h(\bm r)}{h_0} = \sum_{\G} \mathcal V_{\G} \, e^{i \G \cdot \bm r}.
\end{equation}
The Hamiltonian can be diagonalized by Fourier transformation,
\begin{equation}
    \hat \psi_\nu(\bm r) = \frac{1}{\sqrt{V}} \sum_{\bm k} \sum_{\G} e^{i (\bm k - \G ) \cdot \bm r} \hat c_{\nu,\bm k - \G},
\end{equation}
where the sum over $\bm k$ is restricted to the first superlattice Brillouin zone (SBZ) and $\{\hat c_{\nu,\bm k-\G}, \hat c_{\nu,\bm k'-\G'}\} = \delta_{\bm k,\bm k'} \delta_{\G,\G'}$. Note that every wave vector has a unique decomposition as $\bm k - \G$. The Hamiltonian becomes
\begin{equation}
    \hat H_\nu = \frac{1}{V} \sum_{\bm k, \bm k'} \sum_{\G, \G'} \int d^2\bm r \, \hat c_{\nu,\bm k'-\G'}^\dag e^{-i(\bm k' - \G') \cdot \bm r} \left\{ \hbar v \left[ \bm k - \G + \frac{\nu e}{\hbar} \, \bm{\mathcal A}(\bm r) \right] \cdot \left( \nu \sigma_x, \sigma_y \right) + \mathcal V(\bm r) \right\} e^{i(\bm k - \G) \cdot \bm r} \hat c_{\nu,\bm k-\G}.
\end{equation}
Next, we note that
\begin{align}
    \frac{1}{V} \int d^2 \bm r \, e^{-i(\bm k' - \G') \cdot \bm r} \, e^{i(\bm k - \G) \cdot \bm r} & = \delta_{\bm k\bm k'} \delta_{\G\G'}, \\
    \frac{1}{V} \int d^2 \bm r \, e^{-i(\bm k' - \G') \cdot \bm r} \, \bm{\mathcal A}(\bm r) \, e^{i(\bm k - \G) \cdot \bm r} & = \delta_{\bm k\bm k'} \bm{\mathcal A}_{\G-\G'}, \\
    \frac{1}{V} \int d^2 \bm r \, e^{-i(\bm k' - \G') \cdot \bm r} \, \mathcal V(\bm r) \, e^{i(\bm k - \G) \cdot \bm r} & = \delta_{\bm k\bm k'} \mathcal V_{\G-\G'}.
\end{align}
For example, for the triangular height profile, we have
\begin{align}
    \bm{\mathcal A}_{\G} & = \frac{1}{2} \sum_{n=1}^3 \bm{\mathcal A}_n \left( ie^{-2i\phi} \delta_{\G,\G_n} - ie^{2i\phi} \delta_{\G,-\G_n} \right) + \overline{\bm{\mathcal A}}_{\G}, \\
    \mathcal V_{\G} & = \frac{\mathcal V_0}{2} \sum_{n=1}^3 \left( e^{i(\phi + \pi/4)} \delta_{\G,\G_n} + e^{-i(\phi + \pi/4)} \delta_{\G,-\G_n} \right),
\end{align}
where $\overline{\bm{\mathcal A}}_{\G}$ corresponds to higher harmonics that can be gauged away (i.e., they do not contribute to the curl of the pseudo vector potential) and
\begin{equation}
    \bm{\mathcal A}_1 = \frac{\mathcal B_0}{\mathcal G} \begin{pmatrix}
    1 \\ 0 \end{pmatrix}, \qquad \bm{\mathcal A}_{2/3} = \frac{\mathcal B_0}{\mathcal G} \begin{pmatrix} -1/2 \\ \pm \sqrt{3}/2 \end{pmatrix}, \qquad \mathcal V_0 = -eE_z h_0,
\end{equation}
with $\mathcal G=4\pi/\sqrt{3}L$ and $E_z$ the electric field perpendicular to the nominal graphene plane. For computational convenience, it can be preferable to use the gauge
\begin{equation}
    \bm{\mathcal A}(\bm r) = - \hat x \mathcal B_0 \sum_{n=1}^3 \frac{\sin \left( \G_n - 2 \phi \right)}{\mathcal G_{ny}},
\end{equation}
which preserves the symmetries of the system up to a gradient term. The Hamiltonian becomes
\begin{equation}
    \hat H_\nu = \sum_{\bm k} \sum_{\G, \G'} \hat c_{\nu,\bm k-\G'}^\dag \left\{ \hbar v \left[ \left( \bm k - \G \right) \delta_{\G\G'} + \frac{\nu e}{\hbar} \, \bm{\mathcal A}_{\G-\G'} \right] \cdot \left( \nu \sigma_x, \sigma_y \right) + \mathcal V_{\G-\G'} \right\} \hat c_{\nu,\bm k-\G},
\end{equation}
which can be diagonalized numerically by taking a finite number of $\G$ vectors. The number of reciprocal lattice vectors is then increased until the results are converged. The output of this calculation yields the energy bands $\varepsilon_{n\bm k}^\nu$ with eigenvectors $C_{n,\bm k-\G}^\nu$ where $n$ is the band index. Leaving out the valley index, the Bloch wave function becomes
\begin{equation}
    \Psi_{n\bm k}(\bm r) = e^{i\bm k \cdot \bm r} u_{n\bm k}(\bm r), \qquad u_{n\bm k}(\bm r) = \frac{1}{\sqrt{V_c}} \sum_{\G} C_{n,\bm k-\G} \, e^{-i\G\cdot\bm r},
\end{equation}
where $u_{n\bm k}$ are the cell-periodic functions, obeying the periodic gauge condition: $u_{n,\bm k+\G}(\bm r) = e^{-i\G \cdot \bm r} u_{n\bm k}(\bm r)$ and normalization $\langle u_{n\bm k} | u_{m\bm k} \rangle_\text{cell} = \delta_{nm}$. For the calculation of the Berry curvature, we need to evaluate overlaps between the cell-periodic functions at neighboring $\bm k$ points,
\begin{align}
    \langle u_{n\bm k} | u_{n\bm k'} \rangle_\text{cell} & = \frac{1}{V_c} \sum_{\G,\G'} \left( C_{n,\bm k - \G} \right)^* C_{n',\bm k' - \G'} \int_\text{cell} d^2\bm r \, e^{i ( \G - \G' ) \cdot \bm r} \\
    & = \sum_{\G} \left( C_{n,\bm k - \G} \right)^* C_{n',\bm k' - \G}.
\end{align}

If we measure momentum and energy in units of $k_0 = 4\pi/3L$ and $\hbar v k_0$, respectively, the continuum model can be written in terms of two dimensionless parameters,
\begin{equation}
    \frac{L}{l_0}, \qquad \frac{\mathcal V_0}{\hbar v k_0},
\end{equation}
where $l_0 = \sqrt{\hbar / e\mathcal B_0} \propto h_0$ is an effective magnetic length. In the following, we take the experimental values of Ref.\ \cite{Mao2020} ($L = 14~$nm and $\mathcal B_0 = 120~$T) which gives $L/l_0 \approx 6$ and $\hbar v k_0 \approx 172$~meV, and regard $\mathcal V_0$ and $\phi$ as tunable parameters. Because this model has a chiral symmetry under simultaneous reversal of the scalar potential, $\sigma_z \mathcal H_\nu[\mathcal V] \sigma_z = - \mathcal H_\nu[-\mathcal V]$, we only need to consider the valence bands. This model symmetry implies
\begin{equation}
    \varepsilon_{n\bm k}^\nu[-\mathcal V] = -\varepsilon_{-n,\bm k}^\nu[\mathcal V], \qquad \Omega_{n\bm k}^\nu[-\mathcal V] =  \Omega_{-n,\bm k}^\nu[\mathcal V],
\end{equation}
with $n$ a nonzero integer such that $n>0$ ($n<0$) corresponds to conduction (valence) bands.

\subsection{Berry curvature and valley Chern number}

 We numerically calculate the Berry curvature and valley Chern numbers. To this end, we first consider a square plaquette of area $\delta^2$ centered at $\bm k$ with corners: $\bm k_1 = \bm k + \tfrac{\delta}{2}(-1, -1)$, $\bm k_2 = \bm k + \tfrac{\delta}{2}(-1, 1)$, $\bm k_3 = \bm k + \tfrac{\delta}{2}(1, 1)$, and $\bm k_4 = \bm k + \tfrac{\delta}{2}(1, -1)$. For a given band, we then consider the gauge-invariant product
 \begin{equation}
    \langle u_{\bm k_1} | u_{\bm k_2} \rangle \langle u_{\bm k_2} | u_{\bm k_3} \rangle \langle u_{\bm k_3} | u_{\bm k_4} \rangle \langle u_{\bm k_4} | u_{\bm k_1} \rangle = \prod_{m=1}^4 \langle u_{\bm k_m} | u_{\bm k_{m+1}} \rangle,
\end{equation}
where $\bm k_5 = \bm k_1$. One can show that
\begin{equation}
    \Omega_{\bm k} = \lim_{\delta \rightarrow 0} \delta^{-2} \, \arg \prod_{m=1}^4 \langle u_{\bm k_m} | u_{\bm k_{m+1}} \rangle, \qquad \text{Tr} \, g_{\bm k} = - \lim_{\delta \rightarrow 0} \delta^{-2} \, \ln \left| \prod_{m=1}^4 \langle u_{\bm k_m} | u_{\bm k_{m+1}} \rangle  \right|,
\end{equation}
where
\begin{equation}
    g_{\bm k}^{ij} = \text{Re} \left( \langle \partial_i u_{\bm k} | \partial_j u_{\bm k} \rangle \right) + \langle u_{\bm k} | \partial_i u_{\bm k} \rangle \langle u_{\bm k} | \partial_j u_{\bm k} \rangle,
\end{equation}
is the Fubiny-Study quantum metric. For convenience, we use a Bravais grid $(k_1,k_2)$ with $\bm k = k_1 \G_1 + k_2 \G_2$ where
\begin{equation}
    \Omega_{\bm k} = \frac{V_c}{(2\pi)^2} \, F_{12}, \qquad F_{12} = i \left( \langle \partial_1 u_{\bm k} | \partial_2 u_{\bm k} \rangle  - \langle \partial_2 u_{\bm k} | \partial_1 u_{\bm k} \rangle \right), \qquad \mathcal C = \frac{1}{2\pi} \sum_{k_1,k_2} F_{12}.
\end{equation}

\subsection{Phase diagrams}

\begin{figure}
    \centering
    \includegraphics[width=0.75\linewidth]{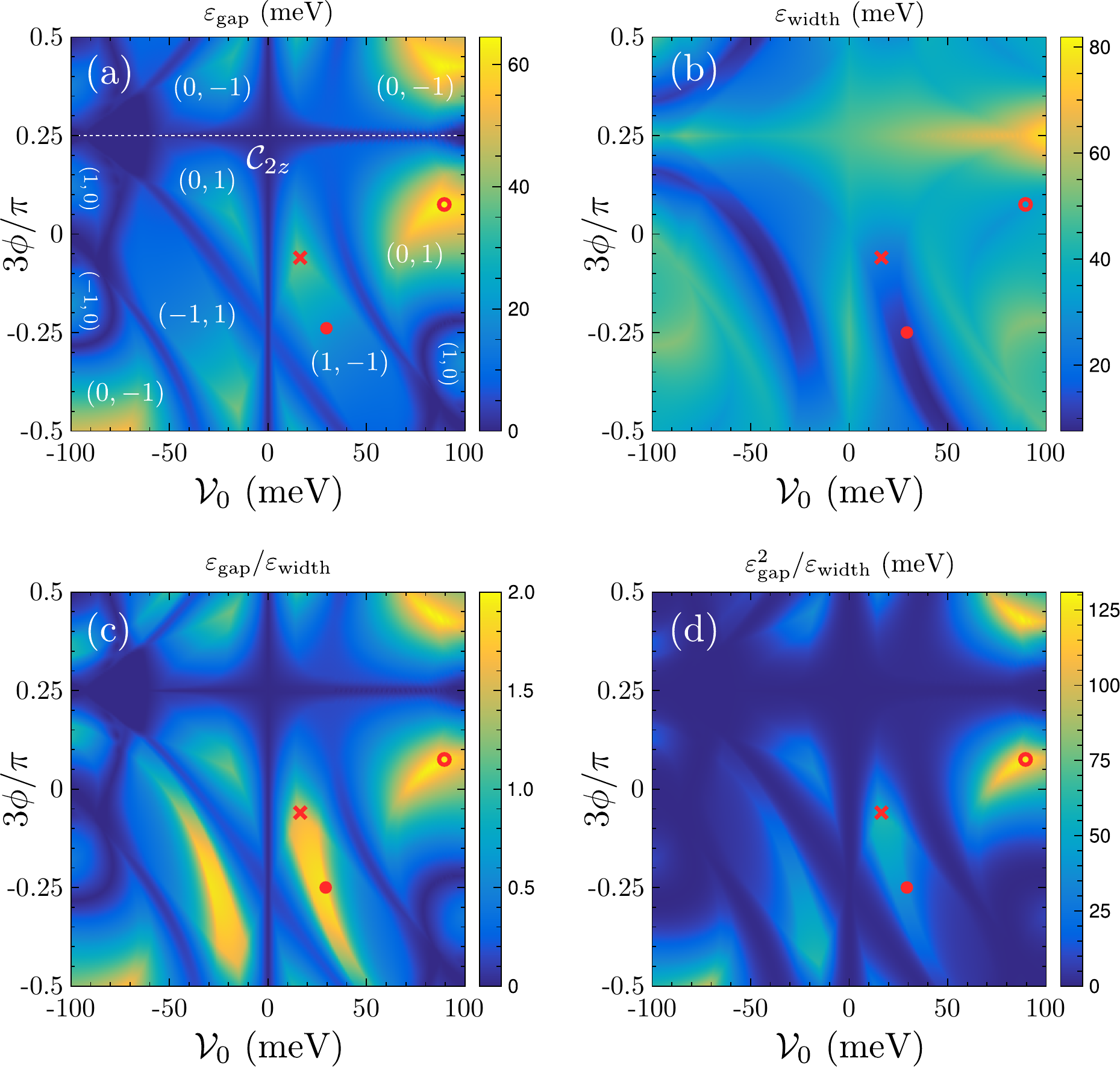}
    \caption{(a) Phase diagram of PBG in the ($\mathcal V_0,\phi)$ plane. The density plot gives the smallest energy gap $\varepsilon_\text{gap}$ of the highest valence band to other bands and the valley Chern numbers of the highest valence and lowest conduction band for valley $K_+$ is shown as $(\mathcal C_{n=-1},\mathcal C_{n=+1})$. (b) Bandwidth $\varepsilon_\text{width}$ of the highest valence band. (c) Ratio $\varepsilon_\text{gap}/\varepsilon_\text{width}$ for the highest valence band. (d) Ratio $\varepsilon_\text{gap}^2/\varepsilon_\text{width}$ for the highest valence band.}
    \label{fig:SI_PBGphase}
\end{figure}

\begin{figure}
    \centering
    \includegraphics[width=0.76\linewidth]{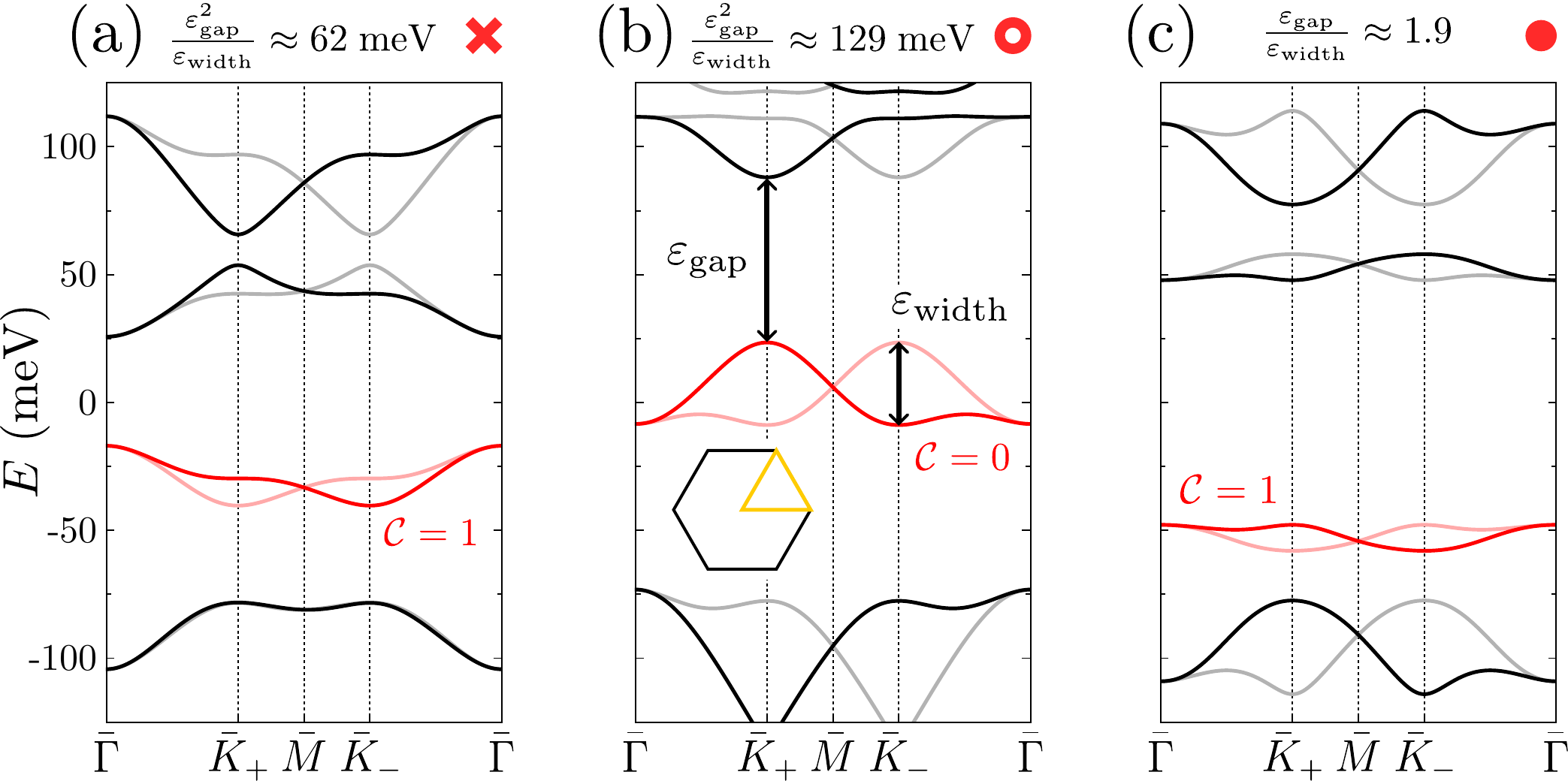}
    \caption{Energy bands of PBG for $L/l_0 = 6$ and $\hbar v k_0 \approx 172$~meV along high-symmetry lines of the SBZ as indicated in (a). The light/dark bands correspond to valley $K_+$/$K_-$ and the red band is the highest valence band. (a) $\mathcal V_0/\hbar v k_0 = 0.095$ and $3\phi/\pi = -0.06$. (b) $\mathcal V_0/\hbar v k_0 = 0.52$ and $3\phi/\pi = 0.075$. (c) $\mathcal V_0/\hbar v k_0 = 0.17$ and $3\phi/\pi = -0.25$.}
    \label{fig:SI_PBGbands}
\end{figure}

We focus on the highest valence band, taking into account both valleys. We numerically calculated the phase diagram in the $(\mathcal V_0,\phi)$ plane for the smallest energy gap to the two neighboring bands $\varepsilon_\text{gap}$. We also calculated the bandwidth $\varepsilon_\text{width}$, as well as the ratios $\varepsilon_\text{gap}/\varepsilon_\text{width}$ and $\varepsilon_\text{gap}^2/\varepsilon_\text{width}$. These diagrams are shown in Fig.\ \ref{fig:SI_PBGphase}. Notice that these diagrams are invariant under $(\mathcal V_0,\phi) \mapsto (-\mathcal V_0,\phi+\pi/3)$. On the phase diagram showing the energy gap, we have indicated the valley Chern numbers of the highest valence band and the lowest conduction band. The energy bands along high-symmetry lines of the SBZ for the parameters corresponding to the circle, disk, and cross in the phase diagrams, are shown in Fig.\ \ref{fig:SI_PBGbands}. 
In the main text, we were mainly interested in showing that the ratio $\varepsilon_\text{gap}^2/\varepsilon_\text{width}$ can be made large enough such that electric breakdown is absent even in the strong-field limit, i.e., the regime
\begin{equation}
     \frac{0.66 \, \text{ps}}{\tau}\frac{10 \, \text{nm}}{L} \ll \frac{E}{\text{kV/cm}} \ll \frac{\varepsilon_\text{gap}^2}{\varepsilon_\text{width} \text{meV}} \frac{10 \, \text{nm}}{L}.
\end{equation}

\section{Twisted double bilayer graphene}

We give an overview of the continuum model for twisted double bilayer graphene, following Ref.\ \cite{Koshino2019}. Before we proceed, we give a short review of the continuum theory of Bernal bilayer graphene.

\subsection{Bernal bilayer graphene}

\begin{figure}
    \centering
    \includegraphics[width=0.95\linewidth]{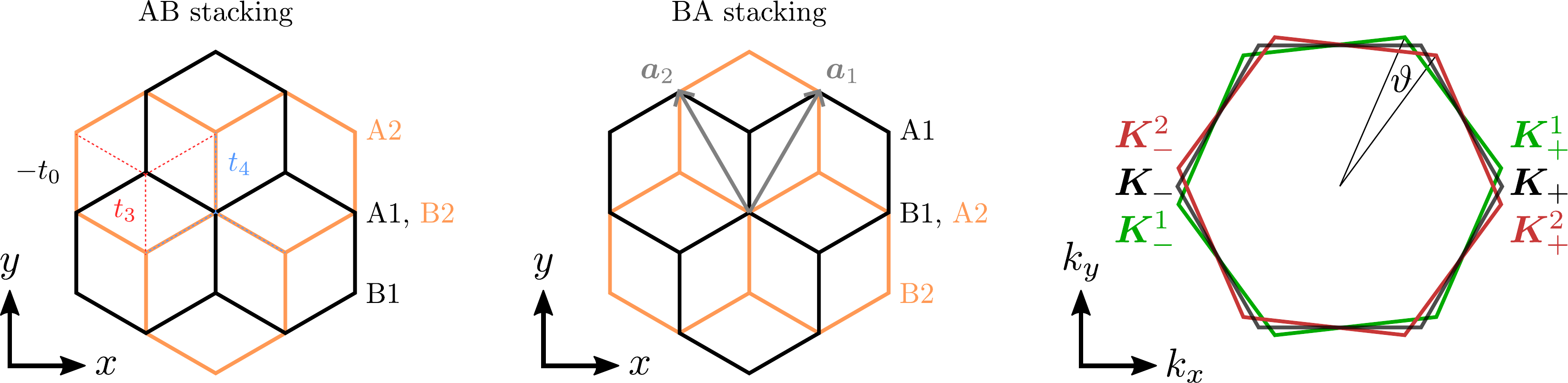}
    \caption{Different stacking configurations of Bernal bilayer graphene and the first BZ of graphene with the two valleys $\bm K_\pm = (\pm4\pi/3a, 0)$ and the rotated zone corners of the two bilayers $\bm K_\pm^1$ and $\bm K_\pm^2$. On the left, we indicate the intralayer nearest-neighbor hopping $-t_0$, and the skew interlayer hoppings $t_3$ (intersublattice) and $t_4$ (intrasublattice) with dashed lines.}
    \label{fig:SI_Bernal}
\end{figure}

We consider a Bernal-stacked bilayer graphene. The sublattices on the first layer are denoted as $A_1$ and $B_1$, and those on the second layer as $A_2$ and $B_2$. We define AB-stacked (BA-stacked) bilayer graphene as the stacking configuration where the atoms of $A_1$ ($B_1$) and $B_2$ ($A_1$) eclipse one another. This is illustrated in Fig.\ \ref{fig:SI_Bernal}.

Following \cite{McCann2013}, we use a lattice model for Bernal bilayer graphene that takes into account intralayer nearest-neighbor hopping with amplitude $-t_0$, interlayer hopping between eclipsing atoms $t_1$, as well as skew interlayer hopping $t_3$ (intersublattice) and $t_4$ (intrasublattice), and a sublattice staggering potential $\delta$. The latter is due to the different environment of the eclipsing atoms. The point group of Bernal bilayer graphene is $D_{3d} = \left< \mathcal C_{3z}, \mathcal C_{2y} , \mathcal I \right>$ where $\mathcal I$ is spatial inversion [$(\bm r,z) \mapsto (-\bm r,-z)$]. Applying an interlayer bias potential breaks inversion symmetry and reduces the point group to $D_3 = \left< \mathcal C_{3z}, \mathcal C_{2y} \right>$.

The corresponding Bloch Hamiltonian of AB Bernal bilayer graphene is given by 
\begin{equation}
h_\text{AB}(\bm k) =
\begin{pmatrix}
    U_1  + \delta & -t_0 f(\bm k)^* & t_4 f(\bm k) & t_1 \\
    -t_0 f(\bm k) & U_1 & t_3 f(\bm k)^* & t_4 f(\bm k) \\
    t_4 f(\bm k)^* & t_3 f(\bm k) & U_2 & -t_0 f(\bm k)^* \\
    t_1 & t_4 f(\bm k)^* & -t_0 f(\bm k) & U_2 + \delta
\end{pmatrix} \equiv \begin{pmatrix}
    h_0(\bm k) + U_1 & g^\dag(\bm k) \\
    g(\bm k) & h_0'(\bm k) + U_2
\end{pmatrix},
\end{equation}
and thus
\begin{equation}
h_\text{BA}(\bm k) = \begin{pmatrix}
    h_0'(\bm k) + U_1 & g(\bm k) \\
    g^\dag(\bm k) & h_0(\bm k) + U_2
\end{pmatrix},
\end{equation}
where
\begin{equation}
    f(\bm k) = 1 + e^{i\bm k \cdot \bm a_1} + e^{i\bm k \cdot \bm a_2},
\end{equation}
and $\bm a_{1/2} = a ( \pm 1/2 ,\sqrt{3}/2)$ with $a \approx 0.246$~nm, see Fig.\ \ref{fig:SI_Bernal}. The sign difference between the intralayer and interlayer hoppings comes from the relative sign of the overlap of $p_z$ orbitals within and between the layers. We take the following values for the hopping constants $t_0 = 2.7$~eV (below we use an effective $t_0$), $t_1 = 0.4$~eV, $t_3 = 0.32$~eV, $t_4 = 0.044$~eV, and $\delta = 0.05$~eV \cite{Koshino2019}. Defining the two valleys as $\bm K_\pm = (\pm 4\pi/3a,0)$, we find
\begin{equation}
    f(\bm k + \bm K_\pm) \simeq -\frac{\sqrt{3}}{2} a \left( \pm k_x + i k_y \right),
\end{equation}
up to first order in $|\bm k|$. Hence, if we place the origin of the momentum at $\bm K_\pm$,
\begin{equation}
    h_{0\nu}(\bm k) \simeq \begin{bmatrix} \delta & \hbar v \left( \nu k_x - i k_y \right) \\ \hbar v \left( \nu k_x + i k_y \right) & 0
    \end{bmatrix}, \qquad h_{0\nu}'(\bm k) \simeq \begin{bmatrix} 0 & \hbar v \left( \nu k_x - i k_y \right) \\ \hbar v \left( \nu k_x + i k_y \right) & \delta
    \end{bmatrix},
\end{equation}
where $\nu = \pm 1$ corresponds to valley $\bm K_\pm$, respectively, and $\hbar v = \sqrt{3}t_0a/2$. In our TDBG calculations, we follow Refs.\ \cite{Moon2013} and \cite{Koshino2019} and take a smaller value of $\hbar v \approx 525.308$~meV~nm, corresponding to an effective $t_0$ of $2.1354$~eV, due to longer-range intersublattice hopping within a single graphene layer in their model. We also have
\begin{equation}
    g_\nu(\bm k) \simeq \begin{bmatrix} -\hbar v_4 \left( \nu k_x - i k_y \right) & -\hbar v_3 \left( \nu k_x + i k_y \right) \\ t_1 & -\hbar v_4 \left( \nu k_x - i k_y \right)
    \end{bmatrix},
\end{equation}
with $\hbar v_3 = \sqrt{3}t_3a/2$ and $\hbar v_4 = \sqrt{3}t_4a/2$.

\subsection{Twisted double bilayer graphene}

We now consider twisted double bilayer graphene (TDBG). Notice that in the absence (presence) of an interlayer bias, the point group of Bernal TDBG is given by $D_3$ ($C_3$) \cite{Koshino2019}. Rotating the Bernal bilayer graphene by an angle $\vartheta$ is equivalent to sending
\begin{equation}
    f(\bm k) \mapsto 1 + e^{i\bm k \cdot R(\vartheta) \bm a_1} + e^{i\bm k \cdot R(\vartheta) \bm a_2} = f(R(-\vartheta)\bm k),
\end{equation}
in the lattice model, with $R(\vartheta)$ the $2\times 2$ rotation matrix. Hence the rotated Dirac points are located at $R(\vartheta)\bm K_\pm$. We construct TDBG by first stacking two Bernal bilayers directly on top of each other, and then rotating the first bilayer by $+\vartheta/2$ and the second bilayer by $-\vartheta/2$. Only the second and third graphene layers are coupled by the interlayer moir\'e potential
\begin{equation}
    T_\nu(\bm r) = T_{\nu 0} + T_{\nu 1} e^{i \nu \G_1 \cdot \bm r} + T_{\nu 2} e^{i \nu \G_2 \cdot \bm r},
\end{equation}
where $\G_{1/2} = ( 4\pi/\sqrt{3}L ) ( \pm 1/2, \sqrt{3}/2)$ are moir\'e reciprocal lattice vectors, and
\begin{equation}
    T_{\nu n} = w_0 \sigma_0 + w_1 \left[ \cos \left( \frac{2\pi n}{3} \right) \sigma_x + \nu \sin \left( \frac{2\pi n}{3} \right) \sigma_y \right],
\end{equation}
with $w_0 = 79.7$~meV and $w_1 = 97.5$~meV the AA and AB interlayer moir\'e amplitudes \cite{Koshino2018,Koshino2019}.

For AB--AB stacked TDBG, we have
\begin{equation}
    \hat H_\nu^\text{AB--AB} = \int d^2 \bm r \, \hat \psi_\nu^\dag(\bm r) \, \mathcal H_\nu^\text{AB--AB}(-i\nabla) \, \hat \psi_\nu(\bm r),
\end{equation}
with $\hat \psi_\nu = ( \hat \psi_{\nu,A1}, \hat \psi_{\nu,B1}, \hat \psi_{\nu,A2}, \hat \psi_{\nu,B2}, \hat \psi_{\nu,A3}, \hat \psi_{\nu,B3}, \hat \psi_{\nu,A4}, \hat \psi_{\nu,B4} )^t$ and
\begin{equation}
\mathcal H_\nu^\text{AB--AB}(-i\nabla) =
\begin{pmatrix}
    h_{0\nu}(\bm k_1) + U_1 & g_\nu^\dag(\bm k_1) & 0 & 0 \\
    g_\nu(\bm k_1) & h_{0\nu}'(\bm k_1) + U_2 & T_\nu^\dag(\bm r) & 0 \\
    0 & T_\nu(\bm r) & h_{0\nu}(\bm k_2) + U_3 & g_\nu^\dag(\bm k_2) \\
    0 & 0 & g_\nu(\bm k_2) & h_{0\nu}'(\bm k_2) + U_4
\end{pmatrix},
\end{equation}
where $\bm k_{1/2} = R(\mp \vartheta/2) \left( -i\nabla - \nu \bm q_{1/2} \right)$ with $\bm q_{1/2} = k_\vartheta (\sqrt{3}/2,\pm 1/2)$ and $k_\vartheta = 4\pi/3L$ with $L = a/2\sin(\vartheta/2)$ the moir\'e lattice constant. Here we have placed the origin of momentum in the center of the moir\'e Brillouin zone (MBZ). We further take $U_1 = U/2$, $U_2=U/6$, $U_3 = -U/6$, and $U_4=-U/2$, such that $U$ is the bias between the topmost and bottommost layer. Similarly, we have for AB--BA stacked TDBG,
\begin{equation}
    \hat H_\nu^\text{AB--BA} = \int d^2 \bm r \, \hat \psi_\nu^\dag(\bm r) \, \mathcal H_\nu^\text{AB--BA}(-i\nabla) \, \hat \psi_\nu(\bm r),
\end{equation}
with
\begin{equation}
\mathcal H_\nu^\text{AB--BA}(-i\nabla) =
\begin{pmatrix}
    h_{0\nu}(\bm k_1) + U_1 & g_\nu^\dag(\bm k_1) & 0 & 0 \\
    g_\nu(\bm k_1) & h_{0\nu}'(\bm k_1) + U_2 & T_\nu^\dag(\bm r) & 0 \\
    0 & T_\nu(\bm r) & h_{0\nu}'(\bm k_2) + U_3 & g_\nu(\bm k_2) \\
    0 & 0 & g_\nu^\dag(\bm k_2) & h_{0\nu}(\bm k_2) + U_4
\end{pmatrix}.
\end{equation}
The Hamiltonian is diagonalized by Fourier transform,
\begin{equation}
    \hat \psi_\nu(\bm r) = \frac{1}{\sqrt{V}} \sum_{\bm k} \sum_{\G} e^{i(\bm k - \G) \cdot \bm r} \hat c_{\nu,\bm k-\G},
\end{equation}
where the sum over $\bm k$ only runs over the MBZ and $\G$ is a reciprocal lattice vector of the moir\'e. Note that every wave vector has a unique decomposition as $\bm k - \G$. We have
\begin{align}
    \frac{1}{V} \int d^2 \bm r \, e^{-i(\bm k' - \G') \cdot \bm r} \, e^{i(\bm k - \G) \cdot \bm r} & = \delta_{\bm k\bm k'} \delta_{\G\G'}, \\
    \frac{1}{V} \int d^2 \bm r \, e^{-i(\bm k' - \G') \cdot \bm r} \, T_\nu(\bm r) \, e^{i(\bm k - \G) \cdot \bm r} & = \delta_{\bm k\bm k'} T_{\nu,\G-\G'},
\end{align}
with
\begin{equation}
    T_{\nu,\G} = T_{\nu 0} \delta_{\G,\bm 0} + T_{\nu 1} \delta_{\G,\nu\G_1} + T_{\nu 2} \delta_{\G,\nu\G_2}.
\end{equation}

\subsection{Phase diagrams}

\begin{figure}
    \centering
    \includegraphics[width=0.75\linewidth]{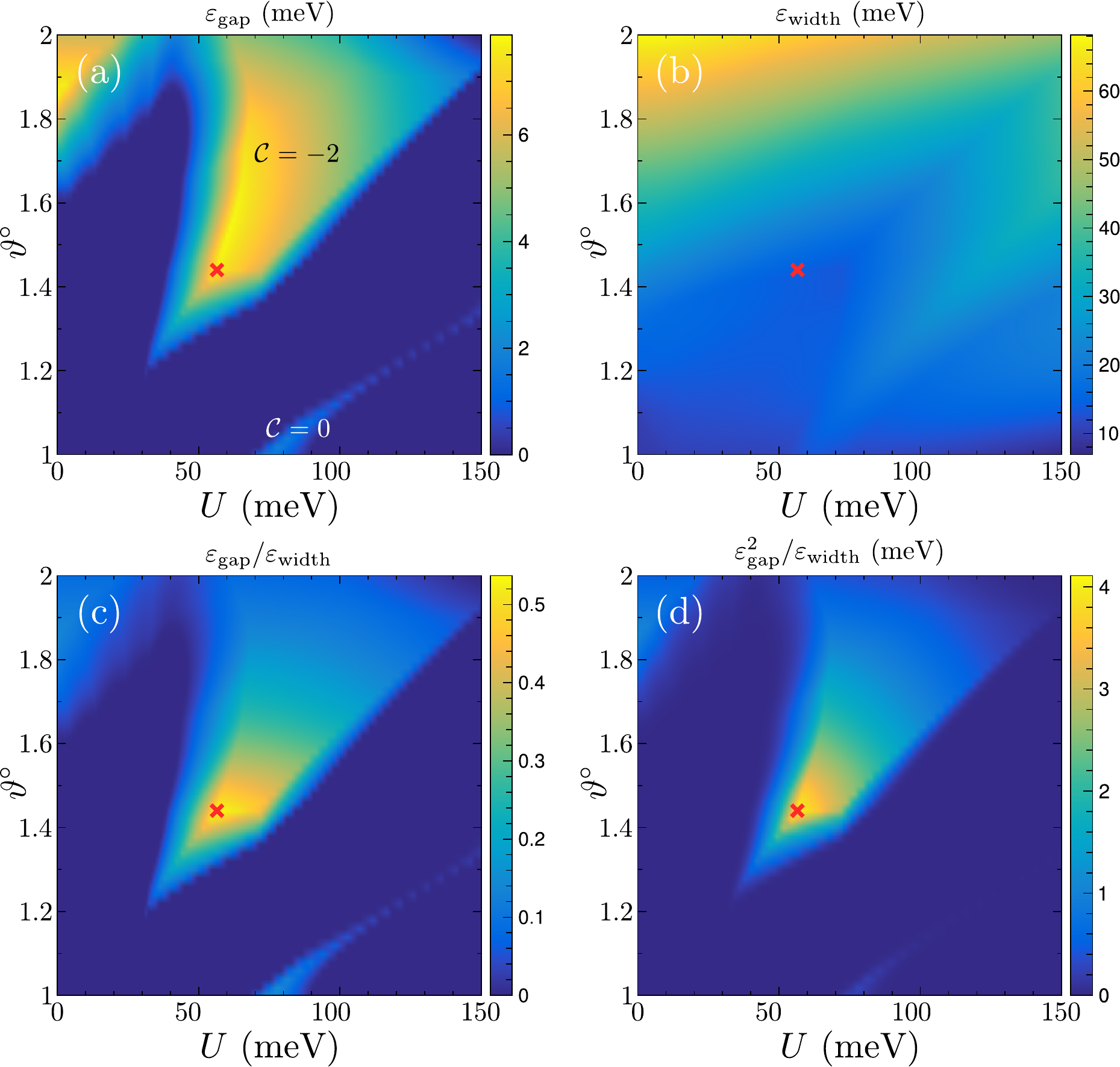}
    \caption{(a) Phase diagram of AB--AB TDBG in the ($U,\vartheta)$ plane. The density plot gives the smallest energy gap $\varepsilon_\text{gap}$ of the lowest conduction band to other bands and the valley Chern number for valley $K_+$. (b) Bandwidth $\varepsilon_\text{width}$ of the lowest conduction band. (c) Ratio $\varepsilon_\text{gap}/\varepsilon_\text{width}$ for the lowest conduction band. (d) Ratio $\varepsilon_\text{gap}^2/\varepsilon_\text{width}$ for the lowest conduction band.}
    \label{fig:SI_TDBG_ABABphase}
\end{figure}

\begin{figure}
    \centering
    \includegraphics[width=0.3\linewidth]{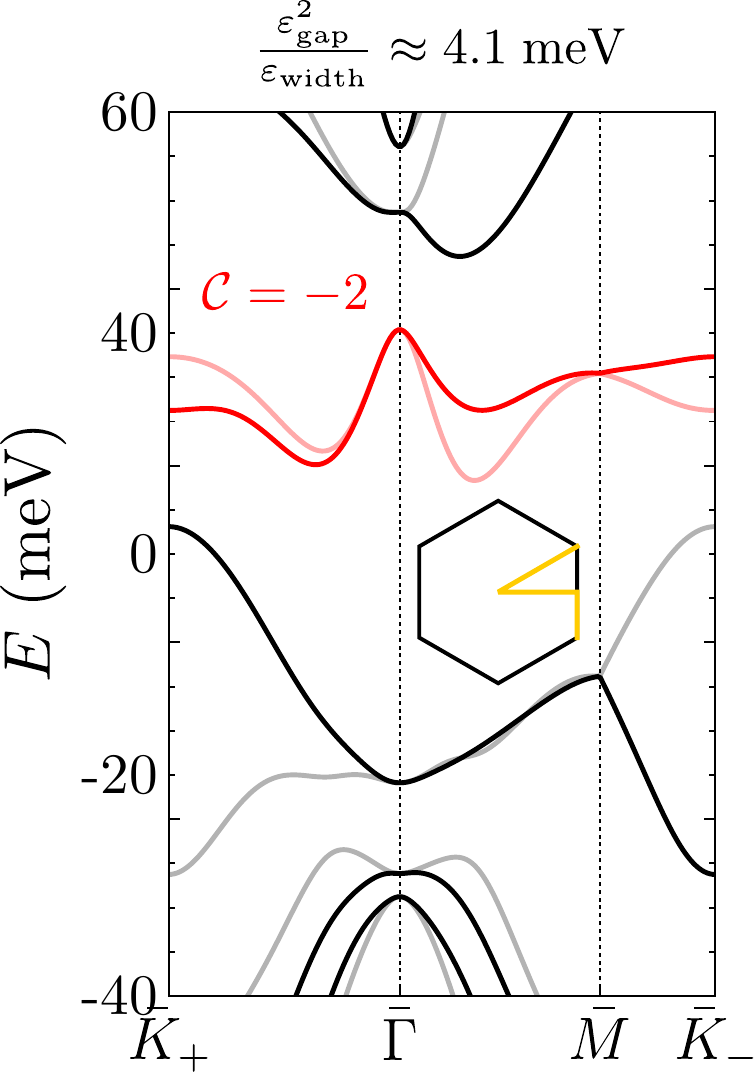}
    \caption{Energy bands of AB--AB TDBG for $\vartheta = 1.44^\circ$ and $U = 56.5$~meV, i.e., the cross in Fig.\ \ref{fig:SI_TDBG_ABABphase} along high-symmetry lines of the SBZ as indicated. Light/dark bands correspond to valley $K_+$/$K_-$ and the red band is the lowest conduction band.}
    \label{fig:SI_TDBG_ABABbands}
\end{figure}

\begin{figure}
    \centering
    \includegraphics[width=0.75\linewidth]{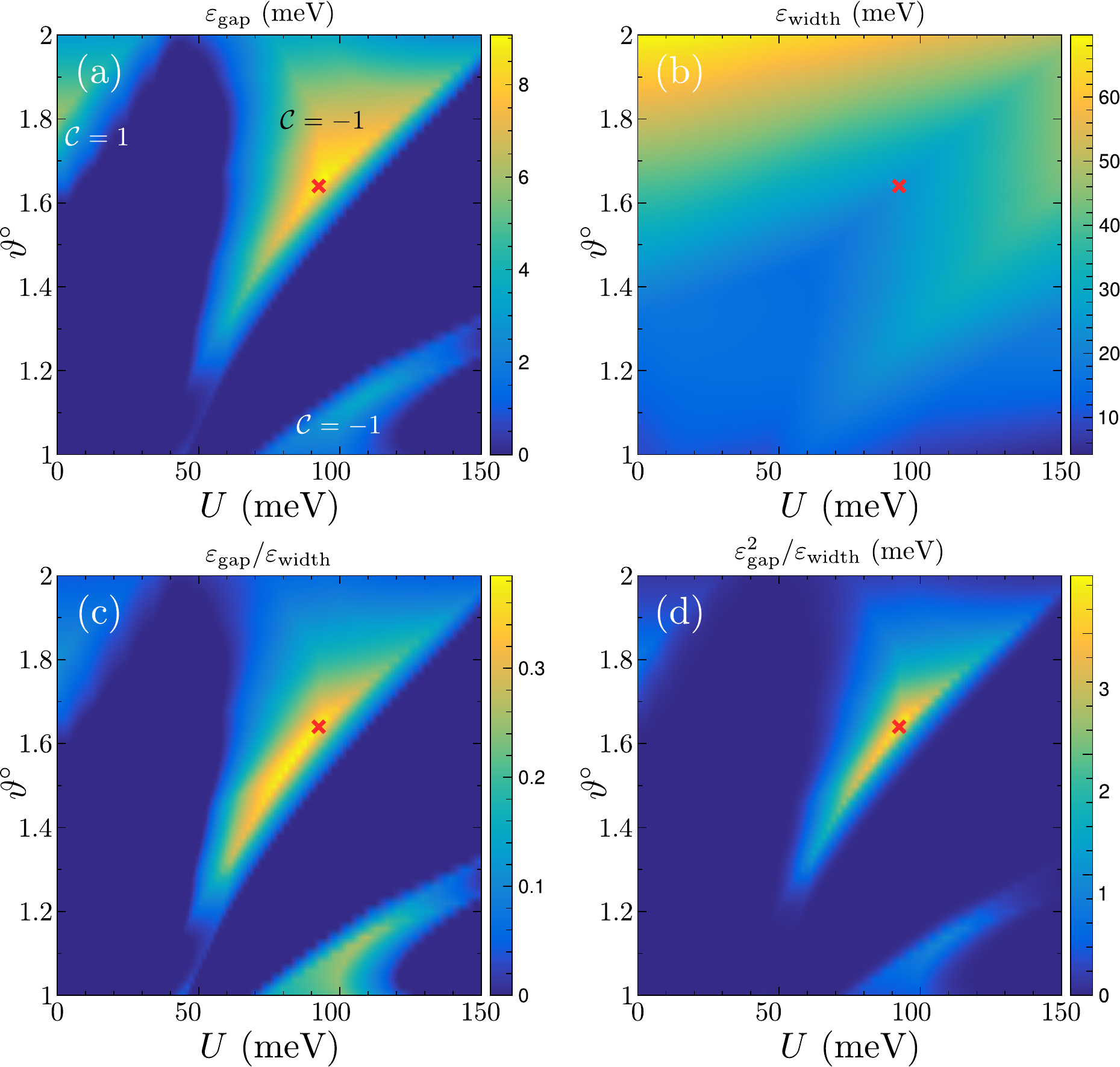}
    \caption{(a) Phase diagram of AB--BA TDBG in the ($U,\vartheta)$ plane. The density plot gives the smallest energy gap $\varepsilon_\text{gap}$ of the lowest conduction band to other bands and the valley Chern number for valley $K_+$. (b) Bandwidth $\varepsilon_\text{width}$ of the lowest conduction band. (c) Ratio $\varepsilon_\text{gap}/\varepsilon_\text{width}$ for the lowest conduction band. (d) Ratio $\varepsilon_\text{gap}^2/\varepsilon_\text{width}$ for the lowest conduction band.}
    \label{fig:SI_TDBG_ABBAphase}
\end{figure}

\begin{figure}
    \centering
    \includegraphics[width=0.3\linewidth]{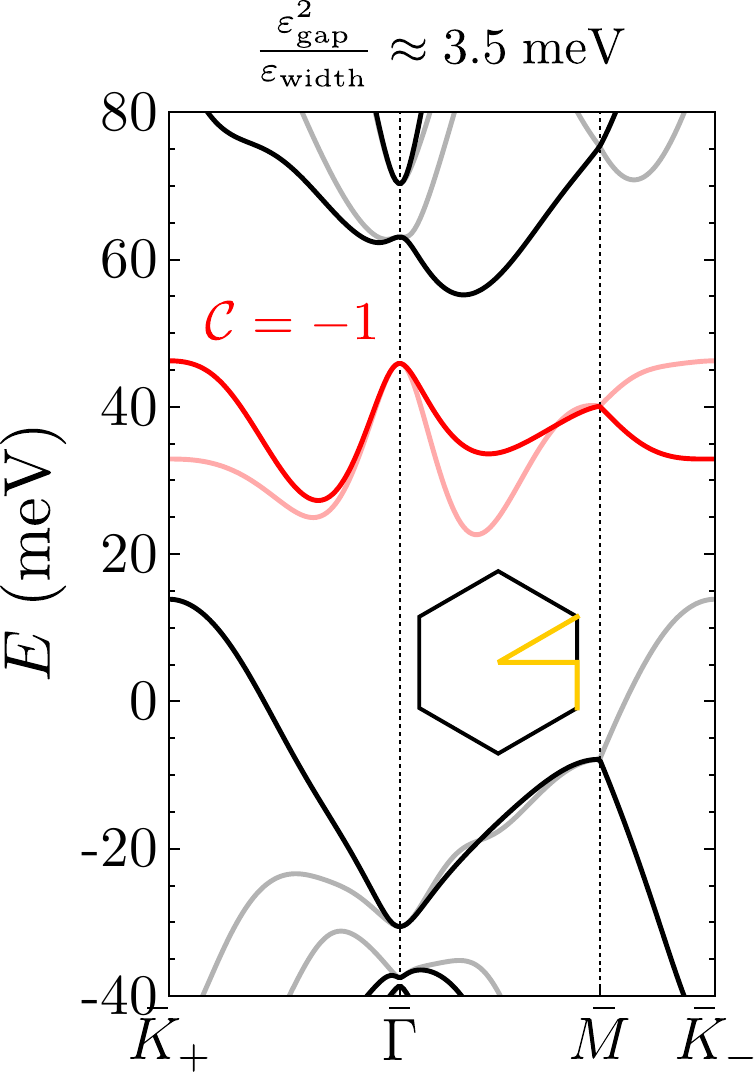}
    \caption{Energy bands of AB--BA TDBG for $\vartheta = 1.64^\circ$ and $U = 92.5$~meV, i.e., the cross in Fig.\ \ref{fig:SI_TDBG_ABBAphase}, along high-symmetry lines of the SBZ as indicated. Light/dark bands correspond to valley $K_+$/$K_-$ and the red band is the lowest conduction band.}
    \label{fig:SI_TDBG_ABBAbands}
\end{figure}

For TDBG we focus on the lowest conduction band. We find that the highest valence band lacks a global energy gap to the remote bands for most of the parameter regime that we considered. We numerically calculated the phase diagram in the $(\vartheta,U)$ plane for the smallest energy gap to the two neighboring bands $\varepsilon_\text{gap}$. We also calculated the bandwidth $\varepsilon_\text{width}$, as well as the ratios $\varepsilon_\text{gap}/\varepsilon_\text{width}$ and $\varepsilon_\text{gap}^2/\varepsilon_\text{width}$. These diagrams are shown in Fig.\ \ref{fig:SI_TDBG_ABABphase} for AB--AB TDBG and in Fig.\ \ref{fig:SI_TDBG_ABBAphase} for AB--BA TDBG, for the lowest conduction band. On the phase diagram showing the energy gap, we have indicated the valley Chern number of the lowest conduction band. The energy bands along high-symmetry lines of the SBZ for the parameters corresponding to the cross in the phase diagrams are shown in Fig.\ \ref{fig:SI_TDBG_ABABbands} for AB--AB TDBG and in Fig.\ \ref{fig:SI_TDBG_ABBAbands} for AB--BA TDBG. 

\end{document}